\documentclass[prl,twocolumn,superscriptaddress,footinbib]{revtex4}

\usepackage[dvipdfmx,hiresbb]{graphicx}
\usepackage{color}
\usepackage{enumerate}

\usepackage{epsfig}
\usepackage{amsmath,amssymb,latexsym}
\usepackage{ascmac}
\usepackage{bm}
\usepackage{enumitem}
\usepackage{natbib}
\def\U#1{{\rm #1}}

\newtheorem{theorem}{{\bf Theorem}}

\newtheorem{pro}{{\bf Proposition}}
\newcommand{\bra}[1]{\langle #1 |}
\newcommand{\ket}[1]{| #1 \rangle}

\newcommand{\expect}[1]{\left\langle #1 \right\rangle}

\newcommand{\sq}{\qquad $\blacksquare$}
\newcommand{\no}{\notag}

\newcommand{\suc}{\U{suc}}
\newcommand{\fin}{\U{final}}
\newcommand{\id}{\U{ideal}}
\newcommand{\sif}{\U{sift}}

\def\Pr{\U{Pr}}
\def\tr{\U{tr}}

\def\vac{{\rm vac}}

\begin{document}
\title
{Security of round-robin differential-phase-shift quantum key distribution protocol with
correlated light sources}
\author{Akihiro Mizutani}
\affiliation{Mitsubishi Electric Corporation, Information Technology R\&D Center,
5-1-1 Ofuna, Kamakura-shi, Kanagawa, 247-8501 Japan}
\author{Go~Kato}
\affiliation{NTT Communication Science Laboratories, NTT Corporation, 3-1,
Morinosato Wakamiya Atsugi-Shi, Kanagawa, 243-0198, Japan}
\begin{abstract}
{
Among various quantum key distribution (QKD) protocols, the round-robin differential-phase-
shift (RRDPS) protocol has a unique feature that its security is guaranteed without monitoring
the signal disturbance. Moreover, this protocol has a remarkable property of being robust against source
imperfections assuming that the emitted pulses are independent. Unfortunately, some experiments with high-speed QKD systems 
confirmed the violation of the independence due to pulse correlations, and therefore the lack of a
security proof with taking into account this effect is an obstacle for guaranteeing the implementation security. 
In this paper, we
show that the RRDPS protocol is secure against any source imperfections by establishing a security proof
with the pulse correlations. The proof is simple in the sense that we make only three experimentally
simple assumptions on the source. Our numerical simulation based on the proof shows that the long-range pulse 
correlation does not cause a significant impact on the key rate, which reveals another
striking feature of the RRDPS protocol. Our security proof is thus effective and applicable to wide
range of practical sources and paves the way to realize truly secure QKD in high-speed systems.
}
\end{abstract}

\maketitle

\section{Introduction}
Quantum key distribution (QKD) offers information-theoretically secure communication
between two distant parties, Alice and Bob~\cite{LoNphoto2014}.
To prove the security of QKD, we suppose mathematical models on the users' devices.
If these models are discrepant from the physical properties of the actual devices, the
security of actual QKD systems cannot be guaranteed.
Hence, it is important to establish a security proof by reflecting the actual properties of
the devices as accurately as possible.

One of the serious imperfections in the source device is the pulse correlation, which
becomes a problem especially in high-speed QKD systems.
Due to experimental imperfections, signal modulation for each emitted pulse affects the modulation of subsequent pulses.
This means that information of Alice's setting choices, such as a bit choice and an intensity choice of the current pulse,
is propagated to the subsequent pulses. 
Indeed, in~\cite{Yoshino2018}, it is experimentally observed that the intensities are correlated among the adjacent pulses
with GHz-clock QKD system. 
Even though tremendous efforts have been made so far to accommodate
imperfections in the source into the security proofs 
(see e.g.~\cite{Diamanti16}), such pulse correlation violates the assumption of most security proofs.
The exceptions are the results in~\cite{Yoshino2018,Zap2021,sciad}, where the 
intensity correlations between the nearest-neighbor pulses and arbitrary intensity 
correlations are respectively accommodated in~\cite{Yoshino2018} 
and~\cite{Zap2021}, and the pulse correlation in terms of Alice's bit choice information 
is taken into account in~\cite{sciad}. 
Note that the result in~\cite{Mizutani2019} provides a security proof incorporating the
correlation among the emitted pulses, but this correlation is assumed to be independent of Alice's setting choices. 

Among various QKD protocols, the round-robin differential-phase-shift (RRDPS) protocol~\cite{rr} is one of the
promising protocols, 
which has a unique feature that its security is guaranteed without monitoring the signal disturbance such as the bit error rate. 
Thanks to this property, the RRDPS protocol has a better tolerance on the bit error rate than the other
protocols and the fast convergence in the finite key regime.
For this protocol, a number of works have been done theoretically
~\cite{MIT2015,rr2016t,rr2017t,sasaki2017,rrt2017H,rr2017tW,Liu2017,Yin2018,Matsuura2019} 
and experimentally~\cite{Takesue2015,Wang2015,Pan2015,Pan2015A,rre2018}. 
Moreover, the RRDPS protocol is shown to be robust against most of source imperfections~\cite{MIT2015}, which is a remarkable property. 
However, this robustness is maintained only when the pulses emitted from the source are independent,
which is also assumed in all the previous security proofs of the RRDPS protocol
~\cite{rr2016t,rr2017t,sasaki2017,rrt2017H,rr2017tW,Liu2017,Yin2018,Matsuura2019}.
Unfortunately, some experiment~\cite{Yoshino2018} confirms the violation of this independence 
due to the pulse correlations, and hence the lack of a security proof with taking into account this effect is an
obstacle for guaranteeing the implementation security of the RRDPS protocol. 

In this paper, we show that the RRDPS protocol is secure against any source imperfections by establishing the 
security proof with the pulse correlations. 
We adopt a general correlation model in which a bit information Alice selected is
encoded
not only on the current pulse but also on the subsequence pulses.
In our security proof, we make only three 
experimentally simple source assumptions, which would be useful for simple source characterization.
More specifically, we assume the length of the correlation among the emitted pulses,
the fidelity between two emitted states when the correlation patterns are different, and
the lower bounds on the vacuum emission probabilities of each emitted pulse.
It is remarkable that no other detailed characterization is required for the source and
any side-channels in the source can be accommodated.
In the security proof, we exploit the reference technique~\cite{sciad} that is a general framework of a
security proof to deal with source imperfections, including the pulse correlation. 
As a result of our security proof, we show that the long-range pulse correlation does
not cause a significant impact on the key rate under a realistic experimental setting, 
which reveals another striking property of the RRDPS protocol.

The paper is organized as follows. 
In section~\ref{sec:idea}, 
we explain how to apply the reference technique to deal with the pulse correlation in the RRDPS protocol and 
why our protocol employs multiple interferometers in Bob's measurement depending on the length of the correlation. 
In sections~\ref{sec:ass} and \ref{sec:pro}, 
we describe the assumptions that we make on Alice and Bob's
devices and introduce the protocol considered, respectively. 
In section~\ref{sec:security}, we first summarize the security proof 
and state our main result about the amount of the privacy amplification, followed by providing its proof.  
Then in section~\ref{sec:simul}, we present our numerical simulation results for the
key generation rate and show that the long-range pulse correlation 
does not cause a significant impact on the key rate. 
Finally, in section~\ref{sec:dis}, we wrap up our security proof 
and refer to some open problems. 

\section{The idea to apply reference technique to RRDPS protocol}
\label{sec:idea}
Here, we explain how to apply the reference technique (RT)~\cite{sciad} to
deal with the pulse correlations in the RRDPS protocol.
In the original RRDPS protocol~\cite{rr}, Alice sends a block of pulses from which Alice
and Bob try to extract one-bit key using a variable-delay interferometer.
On the other hand, in our protocol with the correlation length of $l_c$, Alice and Bob
divide each emitted block into $(l_c+1)$ groups and try to extract $(l_c+1)$-bit key from each of the groups.
In so doing, Bob employs $(l_c+1)$ variable-delay interferometers so that the pulses belonging to the same group interfere.
In other terms, our protocol can be regarded as running $(l_c+1)$ RRDPS protocols
simultaneously.
We adopt such a modification for enabling us to apply the RT. Below, we explain why
the modification is needed.

In the RT, we consider an entanglement-based picture where each $k^{\U{th}}$ emitted pulse is entangled with the qubit.
To discuss the security of the $k^{\U{th}}$ bit $j_k$ that is obtained by measuring the 
qubit in the $Z$-basis (whose eigenstates are denoted by $\{\ket{0},\ket{1}\}$), each qubit is measured in the $X$-basis 
(whose eigenstates are denoted by $\ket{\pm}:=(\ket{0}\pm\ket{1})/\sqrt{2}$).
Since how well Alice can predict the $X$-basis measurement
outcome is directly related to the amount of privacy amplification~\cite{koashi2009},
this estimation is crucial in proving the security.
The RT provides a method for its estimation under the pulse correlation,
but one vital point is that the set of the $k^{\U{th}}$ emitted states must be
fixed just before the emission of the $k^{\U{th}}$ pulse. 
To fix the set, we consider to measure the previous $l_c$ qubits in the $Z$-basis.
For instance, if $l_c=1$, to discuss the security of the even-indexed bit $j_{2k}$,
the previous odd-indexed qubit must be measured in the $Z$-basis.
These $Z$-basis measurements of the previous $l_c$ qubits conflict the original
security proof~\cite{rr} of the RRDPS protocol.
This is because to estimate the aforementioned $X$-basis statistics, all the qubits in the
block are measured in the $X$-basis since any two pulses in the block can interfere in Bob's measurement.
To avoid this conflict, for instance if $l_c=1$, we modify the RRDPS protocol such that
the even-indexed and the odd-indexed pulses interfere separately, and the secret
keys are separately extracted from each interference
using two interferometers.
In doing so, when we discuss the security of the even-indexed bit,
only the even-indexed qubits in the block are measured in the $X$-basis while the odd-indexed
ones are measured in the $Z$-basis.
Hence, thanks to this modification, we can realize both the $X$- and the $Z$-basis
measurements at the same time.
By generalizing this idea to any $l_c\ge2$, if we use $(l_c+1)$ interferometers and
consider the protocol
that extracts the keys from each interferometer, these two basis measurements
become compatible, and hence we can apply
the RT for proving the security.

We remark that when $l_c=1$, the security proofs for the even- and the odd-indexed
keys are mutually exclusive in the sense that the proof for the odd-indexed (even-indexed)
key provides us with how much privacy amplification needs to be applied to
the odd-indexed (even-indexed) key, but it does not offer the security of the even-indexed
(odd-indexed) key. Fortunately, thanks to the universal composability~\cite{composable} of the two
security proofs, the amount of privacy amplification to generate the key both from the
odd- and the even-indexed bits simultaneously is equivalent to those obtained from the
mutually exclusive proofs.
This argument holds for any $l_c\ge2$ due to the universal composability of the $(l_c+1)$ security proofs.

\section{Assumptions on the devices}
\label{sec:ass}
\begin{figure*}[t]
\includegraphics[width=13cm]{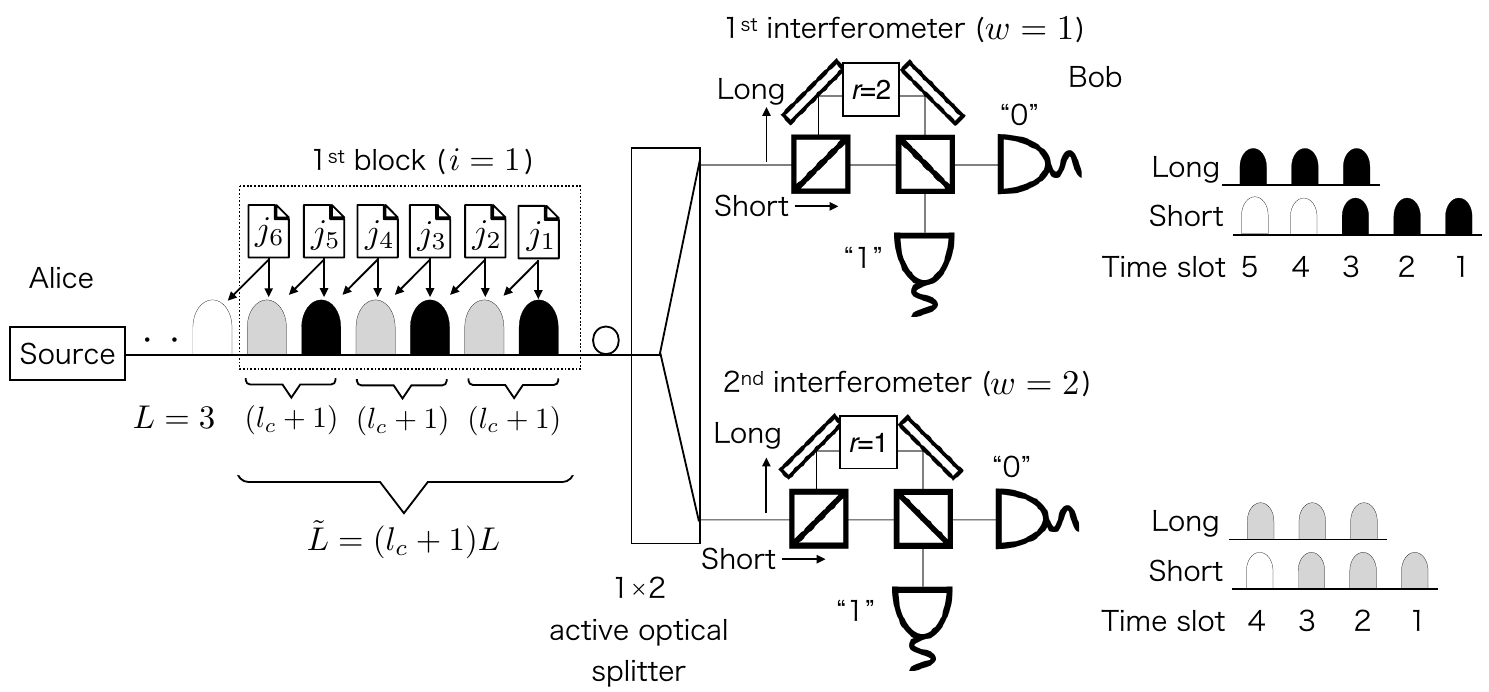}
\caption{
The setups of the source and the measurement devices under nearest-neighbor
correlation ($l_c=1$), 
the number of pulses in each of the block to be fed into each of the interferometer $L$ being three ($L=3$),
and block size being 6 ($\tilde {L}=(l_c+1)L=6$). 
For the $1^{\U{st}}$ block ($i=1$), the positions of the black [gray] pulses belong to 
the $1^{\U{st}}$ group ($w=1$), that is, $\mathcal{G}_{w=1}^{(1)}=\{1,3,5\}$ 
[$2^{\U{nd}}$ group ($w=2$), that is, $\mathcal{G}_{w=2}^{(1)}=\{2,4,6\}$]. 
The black and gray pulses are respectively corresponds to $w=1$ and $w=2$, which are fed
to the $1^{\U{st}}$ and the $2^{\U{nd}}$ variable-delay interferometer with two 50:50 beam splitters, respectively.
Here, the delay $r$ is randomly chosen from the set $\{1,2\}$. 
The pulse trains from the interferometer are measured by two photon-number-resolving
detectors representing bit values ``0" and ``1".
The successful detection in each of the interferometers occurs if Bob detects a single-photon
in total among the $(r+1)^{\U{th}}$ to the $L^{\U{th}}$ time slots and observes no detection at the other time slots.  
}
\label{fig:device}
\end{figure*}
Before describing the protocol, we summarize the assumptions we make on the source
and the receiver.
Figure~\ref{fig:device} depicts the setups of Alice and Bob's devices employed in the
protocol. 
Throughout the paper, we adopt the following notations. 
Let $N$ be the total number of pulses sent by Alice in the protocol, and 
for any symbol $A$, we define $\bm{A}_i:=A_i,A_{i-1},...,A_1$ with $i\in\mathbb{N}$. 

First, we list up the assumptions on Alice's source as follows.
As long as the following assumptions hold, any side-channel in the source can be
accommodated.
\begin{enumerate}[label=(A\arabic*)]
\item
For each $k^{\U{th}}$ emitted pulse ($1\le k\le N$), Alice chooses a random bit
$j_k\in\{0,1\}$.
The bit $j_k$ is encoded not only to the $k^{\U{th}}$ emitted pulse but also to the
subsequence pulses.
Let $l_c\ge0$ be the number of pulses that the information $j_k$ is propagated, and
we call $l_c$ correlation length.
Let 
$\ket{\psi_{j_k|j_{k-1},...,j_1}}_{B_k}=\ket{\psi_{j_k|\bm{j}_{k-1}}}_{B_k}$ be the state of
the $k^{\U{th}}$ emitted signal to Bob, where
the subscripts $j_{k-1},...,j_1$ indicate the dependency of the previous information
$j_{k-1},...,j_1$.
Note that $j_0$ represents having no condition. 
In defining the state $\ket{\psi_{j_k|\bm{j}_{k-1}}}_{B_k}$, we have the freedom in the choice of its global phase. 
Throughout this paper, we fix the global phase of the state $\ket{\psi_{j_k|\bm{j}_{k-1}}}_{B_k}$ 
such that the coefficient of the vacuum state is non-negative. 
In this paper, we consider the case where Alice employs $\tilde{L}$ pulses contained in
a single-block, where $\tilde{L}$ is set to be $(l_c+1)L$ for $L\ge3$.
We call $\tilde{L}$ pulses of systems $B_{(i-1)\tilde{L}+1},...,B_{i\tilde{L}}$ the
$i^{\U{th}}$ block.
\item
When $l_c\ge1$, for any $k$ ($1\le k\le N$) and any $\zeta$ ($k+1\le \zeta\le\min\{N,k+l_c\})$, the following parameter $\epsilon_{\zeta-k}\ge0$
characterizing the correlation is available.
\begin{align}
&\left|\expect{\psi_{j_{\zeta}|j_{\zeta-1},...,j_{k+1},j_k=1,\bm{j}_{k-1}}|\psi_{j_{\zeta}|
j_{\zeta-1},...,j_{k+1},j_k=0,\bm{j}_{k-1}}}\right|^2\no\\
&\ge1-\epsilon_{\zeta-k}.
\label{secII:LRASS1}
\end{align}
Note that the difference between both states in the inner product is in the value of $j_k$. 
The parameter $\epsilon_{\zeta-k}$ depends only on
the difference $\zeta-k$, but it is independent of
$j_{\zeta},j_{\zeta-1},...,j_{k+1},j_{k-1},...,j_1$.
Note that by the assumption (A1),
if $\zeta\ge k+l_c+1$, the left hand size of Eq.~(\ref{secII:LRASS1}) is equal to 1 since
the bit information $j_k$ does not propagate to the $\zeta^{\U{th}}$ state.
\item
For any $k$ $(1\le k\le N)$ and any $j_k\in\{0,1\}$, 
the squared overlap of the vacuum state $\ket{\U{vac}}$ and the state $\ket{\psi_{j_k|j_{k-1},...,j_1}}$ is lower-bounded by
$p^{\U{L}}_{\vac,j_k}$ regardless of $k$ and the previous choices of $j_{k-1},...,j_1$.
Mathematically, we suppose that
\begin{align}
\tr\left[\ket{\vac}\langle{\vac}|\psi_{j_k|\bm{j}_{k-1}}\rangle\bra{\psi_{j_k|\bm{j}
_{k-1}}}\right]
\ge p^{\U{L}}_{\vac,j_k}.
\label{secII:vacP}
\end{align}
\end{enumerate}
Providing the method for experimentally measuring the bounds in 
Eqs.~(\ref{secII:LRASS1}) and (\ref{secII:vacP}) is beyond the scope of this paper.
Note that the assumption~(A2) can be alternatively expressed by using $p^{\U{L}}_{\vac,0}$ 
and $p^{\U{L}}_{\vac,1}$ in Eq.~(\ref{secII:vacP}) because
as we will show in Appendix~\ref{sec:Appe1}, the inner product in Eq.~(\ref{secII:LRASS1}) can be lower-bounded as
\begin{align}
&\left|\expect{\psi_{j_{\zeta}|j_{\zeta-1},...,j_{k+1},j_k=1,\bm{j}_{k-1}}|\psi_{j_{\zeta}|
j_{\zeta-1},...,j_{k+1},j_k=0,\bm{j}_{k-1}}}\right|\no\\
&\ge
\begin{cases}
2p^{\U{L}}_{\vac,j_{\zeta}}-1&\U{if}~2p^{\U{L}}_{\vac,j_{\zeta}}\ge1\\
0&\U{otherwise}.
\end{cases}
\label{eq:onlyvac}
\end{align}

Next, we list up the assumptions on Bob's measurement. 
As explained in Sec.~\ref{sec:idea}, we consider that Alice and Bob try to
extract $(l_c+1)$ secret keys i.e., they divide each block into $(l_c+1)$ groups and try
to generate a one bit key
from each of the groups.
In so doing, Bob employs $(l_c+1)$ variable-delay interferometers with $(L-1)$ delays
followed by two detectors
\footnote{Note that each interferometer followed by two detectors considered in this
paper is the same configuration as the one in the original RRDPS protocol~\cite{rr}. }.
To explain this more clearly,
we classify the set $\{(i-1)\tilde{L}+1,(i-1)\tilde{L}+2,...,i\tilde{L}\}$ of the positions of
the emitted pulses associated with the $i^{\U{th}}$ block into $(l_c+1)$ groups,
and the $w^{\U{th}}$ group ($w\in\{1,2...,l_c+1\}$) for the $i^{\U{th}}$ block is
defined by
\begin{align}
\mathcal{G}_{w}^{(i)}:=\{(l_c+1)(m-1)+w+(i-1)\tilde{L}\}_{m=1}^L.
\label{eq:wth-ith}
\end{align}
Note that $w^{\U{th}}$ group $\mathcal{G}_{w}^{(i)}$ is constructed by picking up all
the $k^{\U{th}}$ pulses from the $i^{\U{th}}$ block with $k\equiv w$ in modulo $
(l_c+1)$.
For instance, if $i=1, l_c=2$, $L=10$ and $\tilde{L}=30$, 
$\mathcal{G}^{(1)}_{w=1}=\{1,4,7,...,28\}, \mathcal{G}^{(1)}_{w=2}=\{2,5,8,...,29\}$ and
$\mathcal{G}^{(1)}_{w=3}=\{3,6,9,...,30\}$.
Then, Bob prepares $(l_c+1)$ interferometers, and for each $i^{\U{th}}$ block, he
feeds the incoming pulses of systems $\{B_k\}_{k\in\mathcal{G}_{w}^{(i)}}$ to the $w^{\U{th}}$ interferometer.
\begin{enumerate}[label=(B\arabic*)]
\item
Bob uses an active optical splitter with one-input and ($l_c+1$)-output to feed the pulses in
the $i^{\U{th}}$ block into the $(l_c+1)$ interferometers. 
This splitter actively sorts the incoming pulses to an appropriate interferometer, where
the $k^{\U{th}}$ pulse with $k\in\mathcal{G}_{w}^{(i)}$ is fed to the $w^{\U{th}}$
interferometer.
\item
Followed by the active optical splitter, Bob employs the $(l_c+1)$ variable-delay
interferometers with two 50:50 beam splitters (BSs),
where the delay of the interferometer is chosen uniformly at random from a set $
\{1,2,...,L-1\}$.
When $r$-bit delay ($r\in\{1,2,...,L-1\}$) is chosen in the interferometer, 
two pulses that are $r(l_c+1)$-pulses apart in terms of the pulses Alice emitted interfere.
\item
After the interferometer, the pulses are detected at time slots 1 through $L+r$ by two
photon-number-resolving (PNR) detectors,
which discriminate the vacuum, a single-photon, and two or more photons of a specific
optical mode.
Each of the detectors is associated to bit values 0 and 1, respectively.
We suppose that the quantum efficiencies and dark countings are the same for both
detectors.
\item
We suppose that there are no side-channels in Bob's measurement device.
\end{enumerate}

\section{Protocol}
\label{sec:pro}
In this section, we describe the actual protocol of the RRDPS protocol under the pulse
correlations in the source device.
Let $N_{\U{em}}$ be the number of emitted blocks sent by Alice, and the total number
of pulses sent by Alice is $N=N_{\U{em}}\tilde{L}$.
As we will see below, our protocol can be regarded as running $(l_c +1)$ RRDPS
protocols simultaneously, each of which employs a block containing
$L$ pulses. More specifically, our protocol runs as follows. 
In the description, $|\bm{z}|$ denotes the length of the bit sequence $\bm{z}$.
\begin{enumerate}
\item
Alice and Bob respectively repeat steps~2 and 3 for $i=1,...,N_{\U{em}}$. \\
\item
Alice chooses a sequence of random bits $j_{(i-1)\tilde{L}+1},...,j_{i\tilde{L}}\in\{0,1\}
^{\tilde{L}}$,
and sends Bob the pulses in the following state through the quantum channel:
\begin{align}
\bigotimes_{k=(i-1)\tilde{L}+1}^{i\tilde{L}}\ket{\psi_{j_k|j_{k-1},...,j_1}}_{B_k}.
\label{eq:emittedActual}
\end{align}
\item
By the active optical splitter with one-input and ($l_c+1$)-output, the pulses in the $i^{\U{th}}
$ block are split to feed into the $(l_c+1)$ variable-delay interferometers.
Among the pulses in the $i^{\U{th}}$ block, the $k^{\U{th}}$ pulse with
$k\in\mathcal{G}_{w}^{(i)}$ is fed to the $w^{\U{th}}$ interferometer.
\\\\
Bob executes the following for $w=1,...,l_c+1$. \\
At the $w^{\U{th}}$ interferometer,
Bob randomly selects the delay $r\in\{1,2,...,L-1\}$, splits $L$ incoming pulses into two
trains of pulses using a 50:50 BS, and shifts backwards only one of the two trains by
$r$.
Recall that the time of a single shift is equal to $(l_c+1)$-times as long as the interval of
the neighboring emitted pulses.
Then, Bob lets each of the first $L-r$ pulses in the shifted train interfere
with each of the last $L-r$ pulses in the other train with the other 50:50 BS,
and detects photons with the two PNR detectors at time slots 1 through $L+r$.
\begin{enumerate}
\item
When Bob detects exactly one photon among the $(r+1)^{\U{th}}$ to the $L^{\U{th}}$
time slots and observes
no detection at the other time slots, he records a sifted key bit $z^{(w)}_{B,i}\in\{0,1\}$
depending on which detector reported the single photon.
He also records the unordered pair $\{u^{(w)}_i,v^{(w)}_i\}$, which are the positions of
the pulse pair
that resulted in the successful detection $(u^{(w)}_i,v^{(w)}_i\in\{1,2,...,L\}, |u^{(w)}_i-v^{(w)}_i|=r)$.
He announces ``success"
and $\{u^{(w)}_i,v^{(w)}_i\}$ over the classical channel.
\item
In all the cases other than (a), Bob announces ``failure" and $w$ through the classical
channel.
\end{enumerate}
\item
Bob executes the following for $w=1,...,l_c+1$. \\
Let $N^{(w)}_{\U{suc}}$ be the number of success blocks observed at the $w^{\U{th}}$
interferometer.
For these blocks,
Bob defines his $w^{\U{th}}$ type sifted key $\bm{z}^{(w)}_{B}$ by concatenating his
sifted key bits $z^{(w)}_{B,i}$ for $i\in\mathcal{B}^{(w)}_{\suc}$.
Here, the set $\mathcal{B}^{(w)}_{\suc}$ is composed of the block-index $i$ where
the pulses whose indices in the set ${\mathcal{G}}^{(i)}_{w}$ result in the successful
detection.
\item
Alice executes the following for $w=1,...,l_c+1$. \\
Alice calculates her sifted key bit $z^{(w)}_{A,i}=j_{k_1}\oplus j_{k_2}$ for $i\in
\mathcal{B}^{(w)}_{\suc}$ with $k_1$ and $k_2$ being the
$u^{(w)}_i$-th and the
$v^{(w)}_i$-th elements of $\mathcal{G}_{w}^{(i)}$, and defines her $w^{\U{th}}$ type
sifted key $\bm{z}^{(w)}_A$
by concatenating her raw key bits $z^{(w)}_{A,i}$ for $i\in\mathcal{B}^{(w)}_{\suc}$.
\item
Bob corrects the bit errors in $\bm{z}_B:=(\bm{z}^{(1)}_B,...,\bm{z}^{(l_c+1)}_B)$
to make it coincide with $\bm{z}_A:=(\bm{z}^{(1)}_A,...,\bm{z}^{(l_c+1)}_A)$ by
sacrificing $|\bm{z}_A|f_{\U{EC}}$ bits of encrypted
public communication from Alice by consuming the same length of a pre-shared secret
key.
\item
Alice and Bob executes the following for $w=1,...,l_c+1$. \\
For each $w^{\U{th}}$ type reconciled key,
Alice and Bob conduct privacy amplification by shortening their keys by $|\bm{z}^{(w)}
_A|f^{(w)}_{\U{PA}}$ to obtain the final keys.
\end{enumerate}
In this paper, we only consider the secret key rate in the asymptotic limit of an infinite
sifted key length. We consider
the asymptotic limit of large $N_{\U{em}}$ while the following observed parameters are
fixed:
\begin{align}
0\le Q^{(w)}:=\frac{N^{(w)}_{\U{suc}}}{N_{\U{em}}}\le1.
\label{eq:Qw}
\end{align}
Note that $f_{\U{EC}}$ in step~6 is determined as a function of the bit error rate
$e_{\U{bit}}$ in $\bm{z}_A$ and $\bm{z}_B$, where
$e_{\U{bit}}$ can be estimated by random sampling whose cost is negligible in the
asymptotic limit.
Also, the fraction of privacy amplification $f^{(w)}_{\U{PA}}$ in step~7
is determined by the experimentally available observables $Q^{(w)}$ in
Eq.~(\ref{eq:Qw}), $\{\epsilon_d\}_{d=1}^{l_c}$ in Eq.~(\ref{secII:LRASS1}), $p^{\U{L}}
_{\vac,0}$ and $p^{\U{L}}_{\vac,1}$ in Eq.~(\ref{secII:vacP}), whose explicit form is
given the next section.

\section{Security proof}
\label{sec:security}

\subsection{Summary of security proof}
\label{sec:securityA}
Here, we summarize the result of the security proof of the protocol described above
and determine the
amount of privacy amplification $|\bm{z}^{(w)}_A|f^{(w)}_{\U{PA}}$ for the $w^{\U{th}}$
type sifted key in the asymptotic limit.
As will be explained in this section,
our security proof is based on the complementarity scenario~\cite{koashi2009}
in which estimation of an upper bound on the phase error rate assures the security.
The main result is this upper bound, which is given in Theorem~\ref{th1}, and we
provide its derivation in Sec.~\ref{securityproof}.
Here, we outline the crux of the discussions.
The difficulty of our phase error rate estimation comes from the correlations among the
emitted pulses
that have not been accommodated in the previous security proofs of the RRDPS protocol
~\cite{MIT2015,rr2016t,rr2017t,sasaki2017,rrt2017H,rr2017tW,Liu2017,Yin2018,Matsuura2019}. 
We solve this problem by exploiting the {\it reference technique} established in~\cite{sciad}.
This is a technique that simplifies the estimation of the phase error rate when the
actually employed states are close to the ones whose formula associated to the phase
error rate is easily derived. In this technique, we consider reference states, which are
fictitious states that are not prepared in the protocol but close to the actual state. The
key intuition is rather simple; when the reference states and the actual states are close,
the deviation between probabilities associated to the reference states and those
associated to the actual states should not be large. Therefore, we can obtain the 
phase error rate formula for the actual states by slightly modifying the
formula for the reference states. We emphasize that Alice does not need to generate
the reference states in the protocol, and they are purely a mathematical tool for phase
error rate estimation.
In particular, we choose the reference states regarding the $k^{\U{th}}$ emitted pulse
such that
the information $j_k$ is {\it only} encoded to system $B_k$ (see Eq.~(\ref{defPHIGC})
for the explicit formula).
By exploiting this property, it is simple to obtain the probabilities for the reference
states,
which will be given by $T$ in Eq.~(\ref{cond1}).
Depending on the fidelity between the actual and reference states, which will be given
by $S$ in Eq.~(\ref{cond2}), by slightly modifying the relationship for the reference states,
we finally obtain the target probability with the actual ones. 

In the rest of this section, we first explain the structure of the security proof, define the
parameters that are needed to present the main result,
and then describe Theorem~\ref{th1}.
For the security proof with complementarity, we consider alternative entanglement-based
procedures for Alice's state preparation at step~2 and calculation of her raw key bit
$z^{(w)}_{A,i}$ at step~5.
These alternative procedures can be employed to prove the security of the actual
protocol because
the states sent to an eavesdropper (Eve), Bob's measurement, and the final key are
identical to those in the actual protocol.
Also, Bob's
public announcement of the unordered pair $\{u^{(w)}_i,v^{(w)}_i\}$ in the actual
protocol is identical to the one in the alternative protocol.
As for Alice's state preparation at step~2, she alternatively prepares $N$ auxiliary qubits in systems
$\bm{A}_N$, which remain at Alice's laboratory during the whole protocol, and the $N$ 
pulses in systems $\bm{B}_N$ to be sent, in the following state
\begin{align}
&\ket{\Psi}_{\bm{A}_N\bm{B}_N}\no\\
&:=\frac{1}{\sqrt{2^N}}
\sum^1_{j_N=0}\cdots\sum^1_{j_1=0}\bigotimes_{k=1}^N\ket{j_k}_{A_k}
e^{\U{i}\theta_{j_k|\bm{j}_{k-1}}}
\ket{\psi_{j_k|\bm{j}_{k-1}}}_{B_k}.
\label{eq:entbased}
\end{align}
Here, the phase factors $e^{\U{i}\theta_{j_k|\bm{j}_{k-1}}}$ 
can be chosen arbitrary because from Eve's perspective, the states of system $B_k$ in
Eqs.~(\ref{eq:emittedActual}) and (\ref{eq:entbased}) are equivalent. 
However, these factors must be adequately chosen to apply the reference technique for each $w^{\U{th}}$ 
type sifted key, which will be explained in Sec.~\ref{app:dec}. 
As for calculation of the sifted key bit $z^{(w)}_{A,i}=j_{k_1}\oplus j_{k_2}$ at step~5,
this bit can be alternatively extracted by applying 
the controlled-not (CNOT) gate (defined on the $Z$-basis) on the $k_1^{\U{th}}$ and
$k_2^{\U{th}}$ auxiliary qubits of systems $A_{k_1}$ and $A_{k_2}$
with the $k_1^{\U{th}}$ one being the control and the $k_2^{\U{th}}$ one being the
target followed by measuring the $k_2^{\U{th}}$ auxiliary qubit in the $Z$-basis to obtain $z^{(w)}_{A,i}$.

In the complementarity scenario, the discussion of the security of the key 
$\bm{z}_{A}^{(w)}$ is equivalent to consider a virtual scenario of how well Alice can predict the
outcome of the measurement complementary to the one to obtain $z_{A,i}^{(w)}$. 
In particular, we take the $X$-basis measurement
as the complementary basis, and we need to quantify how well Alice can predict its outcome $x_{k_2}\in\{+,-\}$ on system $A_{k_2}$.
As for Bob, instead of aiming at learning $z_{A,i}^{(w)}$, he performs the alternative
measurement that determines which of the $k^{\U{th}}$ pulse in the group $\mathcal{G}^{(i)}_w$ contains the single-photon.
This measurement is complementary to the one for obtaining his sifted key bit 
$z_{B,i}^{(w)}$.
With this alternative measurement, Bob announces the pair $\{u^{(w)}_i,v^{(w)}_i\}$
such that the first index $u^{(w)}_i$ corresponds to the location of the single-photon
and the second index $v^{(w)}_i$ is
chosen uniformly at random from the set $\{1,2...,i-1,i+1,i+2...,L\}$
\footnote{
Note that with additional 50\% loss followed by 
this alternative measurement, 
it is equivalent to the actual measurement in Fig.~\ref{fig:device}~\cite{rr}. 
}. 
Hence, in this virtual scenario,
Alice's task is to predict the outcome $x_{k_2}$ where $k_2$ is chosen uniformly at
random from the group $\mathcal{G}^{(i)}_w$ except for $k_1$.
We define the occurrence of {\it phase error} to be the case where Alice fails in her
prediction of the outcome
$x_{k_2}$.
Let $N_{\U{ph}}^{(w)}$ denote the number of phase errors of the $w^{\U{th}}$ type
sifted key among $|\bm{z}_{A}^{(w)}|$ trials.
Suppose that the upper bound $N_{\U{ph}}^{(w),\U{U}}$ on $N_{\U{ph}}^{(w)}$ is
obtained as
a function of the experimentally available observables $Q^{(w)}$ in Eq.~(\ref{eq:Qw}), 
$\{\epsilon_d\}_{d=1}^{l_c}$ in Eq.~(\ref{secII:LRASS1}), $p^{\U{L}}_{\vac,0}$ and $p^{\U{L}}_{\vac,1}$ in Eq.~(\ref{secII:vacP}).
In this case, in the asymptotic limit, a sufficient fraction of privacy amplification is given by~\cite{koashi2009}
\begin{align}
f_{\U{PA}}^{(w)}=h\left(N_{\U{ph}}^{(w),\U{U}}/N^{(w)}_{\U{suc}}\right),
\end{align}
where $h(x)$ is defined by $h(x)=-x\log_2x-(1-x)\log_2(1-x)$ for $0\le x\le 0.5$ and
$h(x)=1$ for $x>0.5$.
Our main result, Theorem~\ref{th1}, derives the upper bound $e^{(w),\U{U}}_{\U{ph}}$
on the phase error rate
$e^{(w)}_{\U{ph}}:=N_{\U{ph}}^{(w)}/N^{(w)}_{\U{suc}}$
with $Q^{(w)}$, $\{\epsilon_d\}_{d=1}^{l_c}$, $p^{\U{L}}_{\vac,0}$ and $p^{\U{L}}_{\vac,1}$ 
(see Sec.~\ref{securityproof} for the proof).
\begin{theorem}
\label{th1}
In the asymptotic limit of large key length of the $w^{\U{th}}$ type sifted key $|\bm{z}^{(w)}_A|$,
the upper bound on the phase error rate for the $w^{\U{th}}$ type sifted key of the RRDPS protocol is given by
\begin{align}
e^{(w),\U{U}}_{\U{ph}}=\sum_{s=0}^{L-2}\frac{1}{L-1}\min\left\{\frac{\nu(L,s,C)}{Q^{(w)}},1\right\}.
\label{eq:resulteph}
\end{align}
Here, function $\nu(L,s,C)$ is defined by
\begin{align}
\nu(L,s,C):=\sum_{y=s+1}^L\binom{L}{y}C^y(1-C)^{L-y},
\no
\end{align}
where $C= g(T,S)$ if $T\le S^2$ and $C=1$ otherwise with
$T:=1-\left(\sqrt{p_{\vac,0}^{\U{L}}}+\sqrt{p_{\vac,1}^{\U{L}}}\right)^2/4$,
$S:=\left(1+\prod_{d=1}^{l_c}\sqrt{1-\epsilon_d}\right)/2$ for $l_c\ge1$ and S=1 for
$l_c=0$,
and $g(x,y):=x+(1-y^2)(1-2x)+2y\sqrt{(1-y^2)x(1-x)}$.
\end{theorem}
We remark that once characterizations of the source device are completed
(i.e., $\{\epsilon_d\}_{d=1}^{l_c}$, $p^{\U{L}}_{\vac,0}$ and $p^{\U{L}}_{\vac,1}$ are
obtained), $C$ becomes a constant.
Theorem~\ref{th1} reveals that as the correlation length $l_c$ gets larger, $\nu(L,s,C)$
in the expression of the phase error rate
in Eq.~(\ref{eq:resulteph}) generally gets larger because $C=g(T,S)$ is a monotonically
decreasing function of $S$ and $S$ generally gets smaller as $l_c$ becomes larger.

Finally, using Theorem~\ref{th1}, the secret key rate per pulse is given by
\begin{align}
R=\sum_{w=1}^{l_c+1}Q^{(w)}\left[1-f_{\U{EC}}-h\left(e^{(w),\U{U}}_{\U{ph}}\right)
\right]/(l_c+1)L,
\label{eq:keyrate}
\end{align}
where we provide its proof in Appendix~\ref{ap:all}. 

\subsection{Proof of the main result}
\label{securityproof}
In this section, we prove our main result, Theorem~\ref{th1}. 

\subsubsection{Derivation of the phase error rate for the $w^{\U{th}}$ type sifted key}
\label{blockmin}
Here, we derive the upper bound
$e^{(w),\U{U}}_{\U{ph}}$ on the phase error rate 
$e^{(w)}_{\U{ph}}:=N_{\U{ph}}^{(w)}/N^{(w)}_{\U{suc}}$ for the $w^{\U{th}}$ type sifted key ($w\in\{1,2,...,l_c+1\}$).
We remark that the following discussions hold for any $w$.
To derive $e^{(w),\U{U}}_{\U{ph}}$,
we consider performing the $X$-basis measurement on system $A_k$ of $\ket{\Psi}_{\bm{A}_N\bm{B}_N}$ in
Eq.~(\ref{eq:entbased}) with $k$ belonging to the $w^{\U{th}}$ group of indices $\bigcup_{i=1}^{N_{\U{em}}}\mathcal{G}^{(i)}_{w}$, 
and we define the total number of the minus $n_{w,-}^{(i)}$ with $1\le i\le N_{\U{em}}$\ obtained through measuring $L$ qubit systems 
$\{A_k\}_{k\in\mathcal{G}_{w}^{(i)}}$. 
Thanks to these $X$-basis measurements, we can classify $N^{(w)}_{\U{suc}}$ successfully detected blocks according to
$n_{w,-}$, which leads to
$N^{(w)}_{\U{ph}}=\sum_{s=0}^LN^{(w)}_{\U{ph},n_{w,-}=s}$. 
Here, $N^{(w)}_{\U{ph},n_{w,-}=s}$ denotes the number of the phase error events for the $w^{\U{th}}$ type sifted key when $n_{w,-}=s$. 
By considering Bob's alternative measurement explained in Sec.~\ref{sec:securityA},
the probability of failing in the prediction of the $X$-basis measurement outcome
(namely, having the occurrence of the phase error)
when $n_{w,-}=s$ is $s/(L-1)$. 
More precisely, the phase error is defined by obtaining the outcome of the minus in the $X$-basis measurement 
on the target qubit, which is randomly chosen with probability $1/(L-1)$~\cite{rr}. 
Since there are $s$-outcomes of the minus when $n_{w,-}=s$, the probability of obtaining the phase error is at most $s/(L-1)$. 
Importantly, the probability $1/(L-1)$ comes from the random choice of the delay at Bob's measurement assumed in~(B2), and 
Eve cannot distort this probability distribution~\cite{rr}. 
With this phase error probability $s/(L-1)$ when $n_{w,-}=s$, the Chernoff bound leads the following for any $\zeta>0$
\begin{align}
e^{(w)}_{\U{ph}}&
:=\frac{N_{\U{ph}}^{(w)}}{N^{(w)}_{\U{suc}}}
=\frac{
\sum_{s=0}^LN^{(w)}_{\U{ph},n_{w,-}=s}}{N^{(w)}_{\U{suc}}}
\no\\
&\le\sum_{s=0}^{L-1}\frac{s}{L-1}
\frac{N^{(w)}_{\U{suc},n_{w,-}=s}}{N^{(w)}_{\U{suc}}}+\zeta+
\frac{N^{(w)}_{\U{suc},n_{w,-}=L}}{N^{(w)}_{\U{suc}}}\no\\
&=\sum_{s=0}^{L-2}\frac{1}{L-1}
\frac{N^{(w)}_{\U{suc},n_{w,-}>s}}{N^{(w)}_{\U{suc}}}+\zeta.
\label{eq:Nph}
\end{align}
To upper-bound $N^{(w)}_{\U{suc},n_{w,-}>s}/N^{(w)}_{\U{suc}}$, whose trivial upper
bound is 1, 
in addition to the successfully detected blocks, Alice measures the total number of the minus $n_{w,-}^{(i)}$ for the non-detected blocks 
(namely, all the $i^\U{th}$ block with $i\notin\mathcal{B}^{(w)}_{\U{suc}}$). 
In doing so, it is obvious to see that the number $N^{(w)}_{\U{suc},n_{w,-}>s}$ of obtaining $n_{w,-}>s$ among
the detected blocks can never be larger than the one $N^{(w)}_{\U{em},n_{w,-}>s}$ among the {\it emitted} blocks
\footnote{
Note that the relation of the numbers $N^{(w)}_{\U{suc},n_{w,-}>s}\le N^{(w)}_{\U{em},n_{w,-}>s}$ just comes from the inclusion relation 
between the emitted blocks and the detected blocks with $n_{w,-}>s$, 
and this relation holds independently of whether the emitted pulses are correlated or not. 
Similar discussions are seen in Eq.~(12) in the original RRDPS protocol~\cite{rr} and in Eq.~(45) in the DPS protocol~\cite{mizu2019dps}. 
}. 
We note that the number $N^{(w)}_{\U{em},n_{w,-}>s}$ of the emitted blocks 
is fixed once Alice prepares the state $\ket{\Psi}_{\bm{A}_N\bm{B}_N}$ in
Eq.~(\ref{eq:entbased}). 
By overestimating $N^{(w)}_{\U{suc},n_{w,-}>s}$ as 
\begin{align}
N^{(w)}_{\U{suc},n_{w,-}>s}\le N^{(w)}_{\U{em},n_{w,-}>s},
\end{align}
Eq.~(\ref{eq:Nph}) results in
\begin{align}
e^{(w)}_{\U{ph}}\le\sum_{s=0}^{L-2}\frac{1}{L-1}
\min\left\{
\frac{N^{(w)}_{\U{em},n_{w,-}>s}}{N^{(w)}_{\U{suc}}},1\right\}+\zeta.
\label{eq:NphEm}
\end{align}
Hence, the remaining task is to derive the upper bound on $N^{(w)}_{\U{em},n_{w,-}>s}$. 
For this, we evaluate 
the upper bound on the probability of obtaining the outcome of the
minus when system $A_k$ of state $\ket{\Psi}_{\bm{A}_N\bm{B}_N}$ is measured in the $X$-basis with  
$k$ belonging to the $m^{\U{th}}$ element ($1\le m\le L$) of set $\mathcal{G}_{w}^{(i)}$. 
Mathematically, the target for computation is the probability $\Pr[x_t=-|\{x_k\}_{k\in\mathcal{P}^{(w)}_{i,m}}]$ with
\begin{align}
t&:=(i-1)\tilde{L}+w+(m-1)(l_c+1),
\notag\\
\mathcal{P}^{(w)}_{i,m}&:=\bigcup_{a=1}^{i}\{k|k\in\mathcal{G}_{w}^{(a)}, k<t\}.
\notag
\end{align}
If 
\begin{align}
\Pr[x_t=-|\{x_k\}_{k\in\mathcal{P}^{(w)}_{i,m}}, j_{t-1},...,j_{t-l_c}]\le C
\label{pAllZ}
\end{align}
with constant $C$
holds for any $j_{t-1},...,j_{t-l_c}\in\{0,1\}^{l_c}$, where $j_k\in\{0,1\}$ denotes the
$Z$-basis measurement outcome on system $A_k$ of Eq.~(\ref{eq:entbased}), 
applying the Bayes rule leads that the target probability is also upper-bounded by $C$
\footnote{
Note that in Eq.~(\ref{jyouC}), $x_t$ and $\{x_k\}_{k\in\mathcal{P}^{(w)}_{i,m}}$ are not independent. 
For example, when $l_c=1$, $\Pr[x_3|x_1]\neq\Pr[x_3]$. 
This is because $x_1$ influences $j_2$ and $j_2$ does $x_3$. 
However,we can remove the condition $x_1$ when conditioned on $j_2$, namely, 
$\Pr[x_3|x_1,j_2]=\Pr[x_3|j_2]$, which will be explained in Sec.~\ref{app:dec}. 
}
:
\begin{align}
\Pr[x_t=-|\{x_k\}_{k\in\mathcal{P}^{(w)}_{i,m}}]\le C.
\label{jyouC}
\end{align}
The derivation of the upper-bound $C$ in Eq.~(\ref{pAllZ}) involves the reference technique established
in~\cite{sciad}, and we explain its detail in Sec.~\ref{app:dec}. 
Note that as mentioned in Sec.~\ref{sec:idea}, to apply the reference technique to estimate the statistics of $x_t$ in Eq.~(\ref{pAllZ}), 
the previous $l_c$ $Z$-basis measurement outcomes $j_{t-1},...,j_{t-l_c}$ must be fixed, 
where these $Z$-basis measurements are possible because we can discuss the security of 
each $w^{\U{th}}$ type sifted key separately thanks to the modification of the RRDPS protocol. 
With Eq.~(\ref{jyouC}) in hand, by considering the binomial trial with success probability $C$,
the total number of the minus $n_{w,-}^{(i)}$ obtained through measuring $L$ systems
$\{A_k\}_{k\in\mathcal{G}_{w}^{(i)}}$ obeys the
following probability distribution when conditioned on the previous outcomes $n_{w,-}^{(i-1)},...,n_{w,-}^{(1)}$:
\begin{align}
&\Pr[n_{w,-}^{(i)}>s|n_{w,-}^{(i-1)},...,n_{w,-}^{(1)}]\le\no\\
&\sum_{y=s+1}^L\binom{L}{y}C^y(1-C)^{L-y}=:\nu(L,s,C).
\label{bound:minus}
\end{align}
Here, $s$ denotes the integer. Once $s$ is fixed, $\nu(L,s,C)$ is constant
independently of the block index $i$ $(1\le i\le N_{\U{em}})$. 
Since the probability of obtaining $n_{w,-}>s$ for any $i^{\U{th}}$ block is upper-bounded by $\nu(L,s,C)$ from Eq.~(\ref{bound:minus}), 
for deriving $N^{(w)}_{\U{em},n_{w,-}>s}$, we can imagine independent trials with probability $\nu(L,s,C)$.
Therefore, again by using the Chernoff bound, we have 
from Eq.~(\ref{eq:NphEm}) and $N^{(w)}_{\U{em},n_{w,-}>s}/N^{(w)}_{\U{suc}}=1/Q^{(w)}\cdot N^{(w)}_{\U{em},n_{w,-}>s}/N_{\U{em}}$
that for any $\chi>0$ and $\zeta>0$
\begin{align}
e^{(w)}_{\U{ph}}\le\sum_{s=0}^{L-2}\frac{1}{L-1}\min\left\{\frac{\nu(L,s,C)+\chi}{Q^{(w)}},1\right\}+\zeta.
\label{finaleph}
\end{align}
When we increase $N^{(w)}_{\suc}$ for any fixed $\zeta$ and $\chi$, the probability of
violating Eq.~(\ref{finaleph})
decreases exponentially.
Therefore, in the limit of large $N^{(w)}_{\suc}$, we can neglect these terms and finally
obtain our main result in Eq.~(\ref{eq:resulteph}).
Note that $Q^{(w)}$ defined in Eq.~(\ref{eq:Qw}) is experimentally observed data.
As will be shown in Sec.~\ref{app:dec}, $C$ is determined by the assumptions (A2) and
(A3), namely, the parameters $\epsilon_1,...,\epsilon_{l_c}$ in Eq.~(\ref{secII:LRASS1}) and the probabilities
$p^{\U{L}}_{\vac,0}$ and $p^{\U{L}}_{\vac,1}$ in Eq.~(\ref{secII:vacP}).

\subsubsection{Derivation of $X$-basis measurement statistics using reference technique}
\label{app:dec}
Here, we derive the upper bound on 
$\Pr[x_t=-|\{x_k\}_{k\in\mathcal{P}^{(w)}_{i,m}},j_{t-1},...,j_{t-l_c}]$ in Eq.~(\ref{pAllZ}),
regardless of $j_{t-1},...,j_{t-l_c}\in\{0,1\}^{l_c}$ that are the $Z$-basis measurement
outcomes of systems $A_{t-1},...,A_{t-l_c}$ in Eq.~(\ref{eq:entbased}).
The crucial point for its computation is that once $j_{t-1},...,j_{t-l_c}$ are fixed, we find
from Eq.~(\ref{eq:entbased}) that the state of the systems
$\bm{A}_{t-1}\bm{B}_{t-1}$ and the one of the systems $\bm{A}_{\ge t}\bm{B}_{\ge t}$ are decoupled, i.e, they are in the tensor product. 
Here, we define $\bm{A}_{\ge i}:=A_N,A_{N-1},...,A_i$. 
The $Z$-basis measurement outcomes $j_{t-1},...,j_{t-l_c}$ have an influence on
determining the set of the $t^{\U{th}}$ states
$\{\ket{\psi_{j_t|\bm{j}_{t-1}}}_{B_t}\}_{j_t}$, but the previous $X$-basis measurement outcomes $\{x_k\}_{k\in\mathcal{P}^{(w)}_{i,m}}$
have no influence on the state of the systems $\bm{A}_{\ge t}\bm{B}_{\ge t}$ thanks
to the tensor product structure.
Therefore, when conditioned on $j_{t-1},...,j_{t-l_c}$, we only focus on the
state of the systems $\bm{A}_{\ge t}\bm{B}_{\ge t}$ to calculate the target
probability.
From Eq.~(\ref{eq:entbased}), conditioned on the outcomes $j_{t-1},...,j_{t-l_c}$, the state of systems $\bm{A}_{\ge t}\bm{B}_{\ge t}$ 
is written as
\begin{align}
\ket{\Gamma^{\U{Act}}_{\bm{j}_{t-1}}}_{\bm{A}_{\ge t}\bm{B}_{\ge t}}:=\frac{1}
{\sqrt{2}}\sum_{j_t=0}^1\ket{j_t}_{A_t}
\ket{\psi^{\U{Act}}_{j_t|\bm{j}_{t-1}}}_{\bm{A}_{\ge t+1}\bm{B}_{\ge t}}
\label{defGNR}
\end{align}
with
\begin{align}
&\ket{\psi^{\U{Act}}_{j_t|\bm{j}_{t-1}}}_{\bm{A}_{\ge t+1}\bm{B}_{\ge t}}:=
e^{\U{i}\theta_{j_t|\bm{j}_{t-1}}}\ket{\psi_{j_t|\bm{j}_{t-1}}}_{B_t}\otimes\no\\
&\frac{1}{\sqrt{2^{N-t}}}\sum_{j_N}\cdots\sum_{j_{t+1}}
\bigotimes_{\zeta=t+1}^N\ket{j_{\zeta}}_{A_{\zeta}}
e^{\U{i}\theta_{j_\zeta|\bm{j}_{\zeta-1}}}
\ket{\psi_{j_{\zeta}|\bm{j}_{\zeta-1}}}_{B_{\zeta}}.
\label{maink-1}
\end{align}
As shown in~\cite{sciad}, and also in Appendix~\ref{sec:ANYCOMP}, Eq.~(\ref{maink-1}) is rewritten as
\begin{align}
&\ket{\psi^{\U{Act}}_{j_t|\bm{j}_{t-1}}}_{\bm{A}_{\ge t+1}\bm{B}_{\ge t}}
=e^{\U{i}\theta_{j_t|\bm{j}_{t-1}}}
\ket{\psi_{j_t|\bm{j}_{t-1}}}_{B_t}\no\\
&\left[
a_{j_t,\bm{j}_{t-1}}\ket{\Phi_{\bm{j}_{t-1}}}_{\bm{A}_{\ge t+1}\bm{B}_{\ge t+1}}
+b_{j_t,\bm{j}_{t-1}}\ket{\Phi^{\perp}_{j_t,\bm{j}_{t-1}}}_{\bm{A}_{\ge t+1}
\bm{B}_{\ge t+1}}\right].
\label{mainstate}
\end{align}
Here, $\ket{\Phi_{\bm{j}_{t-1}}}$ and $\ket{\Phi^{\perp}_{j_t,\bm{j}_{t-1}}}$ are 
normalized states that respectively does not contain the information of $j_t$ and does contain its information.
Note that the state $\ket{\Phi_{\bm{j}_{t-1}}}$ represents a side-channel-free state,
while the state
$\ket{\Phi^{\perp}_{j_t,\bm{j}_{t-1}}}$ represents the state of the side-channel since
the information of $j_t$ is propagated to the subsequence pulses.
In our security proof, $\ket{\Phi^{\perp}_{j_t,\bm{j}_{t-1}}}$ can be taken as any form
in any-dimensional Hilbert space
as long as it is orthogonal to $\ket{\Phi_{\bm{j}_{t-1}}}$, and
the characterization of $\ket{\Phi^{\perp}_{j_t,\bm{j}_{t-1}}}$ is not required. 
As stated below Eq.~(\ref{eq:entbased}), the phase factors $e^{\U{i}\theta_{j_k|\bm{j}_{k-1}}}$ in Eq.~(\ref{eq:entbased}) can be 
chosen arbitrary, but to derive the lower bound on $a_{j_t,\bm{j}_{t-1}}$ in Eq.~(\ref{mainstate}), 
for each $k$ of $k\equiv w$ in modulo $l_c+1$, the phase factors 
$e^{\U{i}\theta_{j_k|\bm{j}_{k-1}}}$ must be set as 
\begin{align}
e^{\U{i}\theta_{j_k|\bm{j}_{k-1}}}=1,
\label{eq:main_ph_1}
\end{align}
and for each $k$ of $k\equiv w$ in modulo $l_c+1$, the phase factors 
$\{e^{\U{i}\theta_{j_\zeta|\bm{j}_{\zeta-1}}} \}_{\zeta=k+1}^{k+l_c}$ must be chosen as 
\begin{align}
&e^{\U{i}\theta_{j_\zeta|\bm{j}_{\zeta-1}}}\no\\
&:=
\frac{\left|\expect{\psi_{j_\zeta|j_{\zeta-1},...,j_{k+1},j_k=0,\bm{j}_{k-1}}|\psi_{j_\zeta|j_{\zeta-1},...,j_{k+1},j_k,\bm{j}_{k-1}}}\right|}
{\expect{\psi_{j_\zeta|j_{\zeta-1},...,j_{k+1},j_k=0,\bm{j}_{k-1}}|\psi_{j_\zeta|j_{\zeta-1},...,j_{k+1},j_k,\bm{j}_{k-1}}}}.
\label{eq:main_ph}
\end{align} 
Note that $e^{\U{i}\theta_{j_t|\bm{j}_{t-1}}}=1$ holds in Eqs.~(\ref{maink-1}) and (\ref{mainstate}) because of Eq.~(\ref{eq:main_ph_1}) and 
$t\equiv w$ in modulo $l_c+1$. 
In doing so, the coefficient $a_{j_t,\bm{j}_{t-1}}$ is positive and can be lower-bounded by using Eq.~(\ref{secII:LRASS1}) as
\begin{align}
a_{j_t=0,\bm{j}_{t-1}}&=1,~~a_{j_t=1,\bm{j}_{t-1}}
\ge \prod_{d=1}^{l_c}\sqrt{1-\epsilon_{d}}
\label{eq:lowa}
\end{align}
if $l_c\ge1$. If $l_c=0$, $a_{j_t,\bm{j}_{t-1}}=1$ for both $j_t=0,1$ (see Appendix~\ref{sec:ANYCOMP} for the detail).

Using Eq.~(\ref{defGNR}), we have that the probability of our interest leads to
\begin{align}
&\Pr[x_t=-|\{x_k\}_{k\in\mathcal{P}^{(w)}_{i,m}}, j_{t-1},...,j_{t-l_c}]\no\\
&=\tr\left[\ket{-}\bra{-}_{A_t}\ket{\Gamma^{\U{Act}}_{\bm{j}_{t-1}}}
\bra{\Gamma^{\U{Act}}_{\bm{j}_{t-1}}}_{\bm{A}_{\ge t}\bm{B}_{\ge t}}
\right].
\label{eq:TC}
\end{align}
To calculate Eq.~(\ref{eq:TC}), we introduce the reference states
$\{\ket{\phi^{\U{Ref}}_{j_t|\bm{j}_{t-1}}}_{\bm{A}_{\ge t+1}\bm{B}_{\ge t}}\}_{j_t}$ 
that are associated with the actual states $\{\ket{\psi^{\U{Act}}_{j_t|\bm{j}_{t-1}}}_{\bm{A}_{\ge t+1}\bm{B}_{\ge t}}\}_{j_t}$.
The reference states, which are close to the actual states prepared by the
protocol, need to be chosen such that the following two conditions are satisfied. In its description, we use the notation
\begin{align}
\ket{\Gamma^{\U{Ref}}_{\bm{j}_{t-1}}}_{\bm{A}_{\ge t}\bm{B}_{\ge t}}:=\frac{1}
{\sqrt{2}}\sum_{j_t=0}^1\ket{j_t}_{A_t}
\ket{\phi^{\U{Ref}}_{j_t|\bm{j}_{t-1}}}_{\bm{A}_{\ge t+1}\bm{B}_{\ge t}}.
\label{eq:RS}
\end{align}
\begin{enumerate}[label=(C\arabic*)]
\item
For the reference state,
the probability of obtaining the outcome of the minus when system $A_t$ is measured
in the $X$-basis is upper-bounded by constant $T>0$, which is expressed as
\begin{align}
\Pr\left[x_t=-~|~\ket{\Gamma^{\U{Ref}}_{\bm{j}_{t-1}}}\right]\le T.
\label{cond1}
\end{align}
\item
The fidelity between $\ket{\Gamma^{\U{Act}}_{\bm{j}_{t-1}}}$ and $
\ket{\Gamma^{\U{Ref}}_{\bm{j}_{t-1}}}$ is lower-bounded by constant $S>0$,
that is
\begin{align}
\left|\expect{\Gamma^{\U{Ref}}_{\bm{j}_{t-1}}|\Gamma^{\U{Act}}_{\bm{j}_{t-1}}}\right|\ge S.
\label{cond2}
\end{align}
\end{enumerate}
Once the reference states satisfy Eqs.~(\ref{cond1}) and (\ref{cond2}), 
the upper bound on Eq.~(\ref{eq:TC}) can be obtained by using the function $g(x,y)$~\cite{sciad} 
that relates the statistics of the actual and the reference states
\footnote{
Note that the exact statement presented in~\cite{sciad} is that for any two normalized
states $\ket{A}$ and $\ket{R}$ and
any POVM (positive-operator-valued measure) $\{M, I-M\}$,
$$
\tr[\ket{A}\bra{A}M]\le g(\tr[\ket{R}\bra{R}M], |\expect{A|R}|).
$$
}.
Specifically, the $X$-basis measurement statistics of these two states are related as
\begin{align}
&\Pr\left[x_t=-~|~\ket{\Gamma^{\U{Act}}_{\bm{j}_{t-1}}}\right]
\no\\
&\le g\left(\Pr\left[x_t=-~|~\ket{\Gamma^{\U{Ref}}_{\bm{j}_{t-1}}}\right],
\left|\expect{\Gamma^{\U{Ref}}_{\bm{j}_{t-1}}|\Gamma^{\U{Act}}_{\bm{j}_{t-1}}}
\right|\right),
\label{PRGAMMA}
\end{align}
where $g(x,y)=x+(1-y^2)(1-2x)+2y\sqrt{(1-y^2)x(1-x)}$ if $x\le y^2$ and $g(x,y)=1$ if $x> y^2$.
A direct calculation reveals that if $x\le y^2$, 
\begin{align}
g(x,y)\le g(x^{\U{U}},y^{\U{L}})
\label{eq:gUL}
\end{align}
holds, where U (L) indicates the upper (lower) bound. Then, combining Eqs.~(\ref{cond1})-(\ref{eq:gUL}) gives

\begin{eqnarray}
\Pr\left[x_t=-~|~\ket{\Gamma^{\U{Act}}_{\bm{j}_{t-1}}}\right] \le
C=
\begin{cases}
g(T,S)& (\U{if}~T\le S^2) \\
1 & (\U{if}~T> S^2).
\end{cases}
\label{eq:fe}
\end{eqnarray}
Hence, the remaining task for obtaining Eq.~(\ref{jyouC}) is to derive the two bounds $T$ and $S$, which are calculated below.
In so doing, we take the reference state
$\ket{\phi^{\U{Ref}}_{j_t|\bm{j}_{t-1}}}_{\bm{A}_{\ge t+1}\bm{B}_{\ge t}}$ for
$j_t\in\{0,1\}$, which are associated with the actual state
$\ket{\psi^{\U{Act}}_{j_t|\bm{j}_{t-1}}}_{\bm{A}_{\ge t+1}\bm{B}_{\ge t}}$ in
Eq.~(\ref{mainstate}), such that it is
the first term of $\ket{\psi^{\U{Act}}_{j_t|\bm{j}_{t-1}}}_{\bm{A}_{\ge t+1}\bm{B}_{\ge t}}$:
\begin{align}
\ket{\phi^{\U{Ref}}_{j_t|\bm{j}_{t-1}}}_{\bm{A}_{\ge t+1}\bm{B}_{\ge t}}=
\ket{\psi_{j_t|\bm{j}_{t-1}}}_{B_t}\otimes\ket{\Phi_{\bm{j}_{t-1}}}_{\bm{A}_{\ge t+1}\bm{B}_{\ge t+1}}.
\label{defPHIGC}
\end{align}
\\
{\bf Calculation of $T$ in Eq.~(\ref{cond1}):}\\
We calculate the upper bound on $\Pr\left[x_t=-~|~\ket{\Gamma^{\U{Ref}}_{\bm{j}
_{t-1}}}\right]$ as follows:
\begin{align}
&\Pr\left[x_t=-~|~\ket{\Gamma^{\U{Ref}}_{\bm{j}_{t-1}}}\right]=1-\Pr\left[x_t=+~|~\ket{\Gamma^{\U{Ref}}_{\bm{j}_{t-1}}}\right]
\notag\\
&=1-\sum_{n=0}^{\infty}\Pr\left[n_t=n, x_t=+~|~\ket{\Gamma^{\U{Ref}}_{\bm{j}_{t-1}}}\right]\notag\\
&\le1-\Pr\left[n_t=0, x_t=+~|~\ket{\Gamma^{\U{Ref}}_{\bm{j}_{t-1}}}\right],
\notag
\end{align}
where $n_t$ denotes the number of photons contained in system $B_t$. 
By rewriting $\ket{\Gamma^{\U{Ref}}_{\bm{j}_{t-1}}}$ using the $X$-basis states $\ket{\pm}_{A_t}$:
\begin{align}
&\ket{\Gamma^{\U{Ref}}_{\bm{j}_{t-1}}}_{\bm{A}_{\ge t}\bm{B}_{\ge t}}=
\ket{\Phi_{\bm{j}_{t-1}}}_{\bm{A}_{\ge t+1}\bm{B}_{\ge t+1}}\otimes\no\\
&\frac{\ket{+}_{A_t}\sum_{j_t}\ket{\psi_{j_t|\bm{j}_{t-1}}}_{B_t}+\ket{-}_{A_t}
\sum_{j_t}(-1)^{j_t}\ket{\psi_{j_t|\bm{j}_{t-1}}}_{B_t}}{2},
\label{eq:XREF}
\end{align}
we find that the statistics of $n_t$ only depends on system $A_t$.
Importantly, in obtaining Eq.~(\ref{eq:XREF}), we used the fact that $\ket{\Phi_{\bm{j}_{t-1}}}_{\bm{A}_{\ge t+1}\bm{B}_{\ge t+1}}$ 
is independent of $j_t$ as stated in Sec.~\ref{app:dec}. Then, combining Eq.~(\ref{eq:XREF}) and the assumption~(A3) in
Sec.~\ref{sec:ass} gives the lower-bound on $\Pr\left[n_t=0, x_t=+~|~\ket{\Gamma^{\U{Ref}}_{\bm{j}_{t-1}}}\right]$ as
\begin{align}
&\Pr\left[n_t=0, x_t=+~|~\ket{\Gamma^{\U{Ref}}_{\bm{j}_{t-1}}}\right]
=\left|\frac{\bra{\vac}\sum_{j_t}\ket{\psi_{j_t|\bm{j}_{t-1}}}}{2}\right|^2\no\\
&\ge\left[\sqrt{p_{\vac,0}^{\U{L}}}+\sqrt{p_{\vac,1}^{\U{L}}}\right]^2/4.
\notag
\end{align}
In the inequality, 
we employ the fact that the coefficient of the vacuum state of $\ket{\psi_{j_t|\bm{j}_{t-1}}}$ 
is non-negative, which is stated in assumption~(A1). 
Therefore,
\begin{align}
\Pr\left[x_t=-~|~\ket{\Gamma^{\U{Ref}}_{\bm{j}_{t-1}}}\right]
\le1-\left(\sqrt{p_{\vac,0}^{\U{L}}}+\sqrt{p_{\vac,1}^{\U{L}}}\right)^2/4=T.
\no
\end{align}
\\
{\bf Calculation of $S$ in Eq.~(\ref{cond2}):}\\
Next, we calculate the fidelity between
$\ket{\Gamma^{\U{Act}}_{\bm{j}_{t-1}}}$ and $\ket{\Gamma^{\U{Ref}}_{\bm{j}_{t-1}}}$. 
We have that for $l_c\ge1$
\begin{align}
&\left|\expect{\Gamma^{\U{Ref}}_{\bm{j}_{t-1}}|\Gamma^{\U{Act}}_{\bm{j}_{t-1}}}
\right|
=
\frac{\left|\sum_{j_t}\expect{\phi^{\U{Ref}}_{j_t|\bm{j}_{t-1}}|\psi^{\U{Act}}_{j_t|\bm{j}_{t-1}}}\right|}{2}
\no\\
&=\frac{\left|\sum_{j_t}a_{j_t,\bm{j}_{t-1}}\right|}{2}
\ge
\frac{1+\prod_{d=1}^{l_c}\sqrt{1-\epsilon_d}}{2}=S.
\notag
\end{align}
The first equality follows from Eqs.~(\ref{defGNR}) and (\ref{eq:RS}), 
the second equality comes from Eqs.~(\ref{mainstate}) and (\ref{defPHIGC}), and the inequality follows from Eq.~(\ref{eq:lowa}).
If $l_c=0$, $S=1$ holds since $a_{j_t,\bm{j}_{t-1}}=1$ for both $j_t=0,1$.

\section{Simulation of secure key rates}
\label{sec:simul}
\begin{figure}[t]
\includegraphics[width=8.5cm]{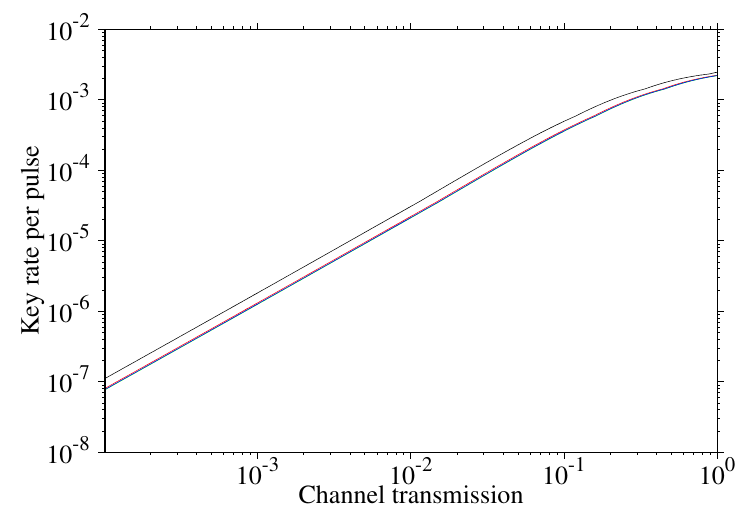}
\caption{
Secure key rate $R$ per pulse as a function of the overall channel transmission $\eta$.
From top to bottom, we plot the key rates for $(l_c,\Delta)=$(0,0), (1,0.2), (2,0.2) and
(10,0.2)
with $e_{\U{bit}}=3\%$ and $L=32$. We note that the key rates for the last three
parameters are almost superposed.
}
\label{fig:rate}
\end{figure}
Here, we show the simulation results of asymptotic key rate $R$ per pulse given by
Eq.~(\ref{eq:keyrate}) as a function of the overall channel
transmission $\eta$ including the detector efficiency. For the simulation, we assume
that each emitted
pulse is a coherent pulse from a conventional laser with mean photon number $\mu$
and only the phases of the coherent pulses are correlated.
In this case, the lower bound on the vacuum emission probability $p^{\U{L}}_{\U{vac},j_k}$ in 
Eq.~(\ref{secII:vacP}) is given by $e^{-\mu}$ and hence 
$T$ defined above is $1-e^{-\mu}$. 
For the simulation, we consider the cases of the correlation length $l_c=0,1,2$ and 10, and for all the cases, 
we adopt $f_{\U{EC}}=h(e_{\U{bit}})$ with $e_{\U{bit}}$ denoting the bit error rate in the protocol and suppose the successful detection rate
for any $w$ as $Q^{(w)}=L\eta\mu e^{-L\eta\mu}/2$.

In the case of $l_c=1$, namely, the case of the nearest-neighbor correlation, we assume
that the $k^{\U{th}}$ emitted state is written as
\begin{align}
\ket{\psi_{j_k|j_{k-1}}}=
\delta_{j_{k-1},0}\ket{(-1)^{j_k}\sqrt{\mu}}
+\delta_{j_{k-1},1}\ket{(-1)^{j_k}\sqrt{\mu} e^{\U{i}\Delta}}.
\label{eq:nn}
\end{align}
Here, $\delta_{x,y}$ is the Kronecker delta and $\ket{e^{\U{i}\theta}\sqrt{\mu}}$
denotes the coherent state with the complex amplitude being $e^{\U{i}\theta}
\sqrt{\mu}$.
This state represents the pulse correlation that if the previous choice of bit $j_{k-1}$ is
0, then the next $k^{\U{th}}$ states are the
ideal states $\{\ket{\sqrt{\mu}},\ket{-\sqrt{\mu}}\}$, but if $j_{k-1}$ is 1, then the
phases are deviated by $\Delta$
from the ideal ones. In this setting, $S$ defined above is given by
\begin{align}
S=\frac{1+\sqrt{1-\epsilon_1}}{2}=\frac{1+|\expect{\sqrt{\mu}|\sqrt{\mu} 
e^{\U{i}\Delta}}|}{2}=\frac{1+e^{\mu(\cos\Delta-1)}}{2}.
\no
\end{align}
In the case of $l_c=2$, we assume that the $k^{\U{th}}$ emitted state is written as
$\ket{\psi_{j_k|j_{k-1},j_{k-2}}}=\delta_{j_{k-1},0}\delta_{j_{k-2},0}\ket{(-1)^{j_k}
\sqrt{\mu}}+
\delta_{j_{k-1},0}\delta_{j_{k-2},1}\ket{(-1)^{j_k}\sqrt{\mu}e^{\U{i}\Delta/2}}
+\delta_{j_{k-1},1}\delta_{j_{k-2},0}\ket{(-1)^{j_k}\sqrt{\mu}e^{\U{i}\Delta}}+
\delta_{j_{k-1},1}\delta_{j_{k-2},1}\ket{(-1)^{j_k}\sqrt{\mu}e^{\U{i}3\Delta/2}}$. 
This state represents that if $j_{k-1}=1$ ($j_{k-2}=1$), the phases of the $k^{\U{th}}$
pulses are rotated by $\Delta$ ($\Delta/2$).
This means that the influence of the second-previous bit $j_{k-2}$ to the $k^{\U{th}}$ pulse
is half of that of the previous bit $j_{k-1}$.
In this setting, $S$ is given by
\begin{align}
S=\frac{1+\sqrt{1-\epsilon_1}\sqrt{1-\epsilon_2}}{2}
=\frac{1+e^{\mu(\cos\Delta-1)}e^{\mu(\cos\frac{\Delta}{2}-1)}}{2}.
\no
\end{align}
As for the case of $l_c=10$, the $k^{\U{th}}$ state is set to be analogous to the one
for $l_c=1,2$, where if $j_{k-d}=1$ (with $d=1,...,10$), the phases of the
$k^{\U{th}}$ pulses are rotated by $\Delta/2^{d-1}$.
A direct calculation shows
$S=\left(1+\prod_{d=1}^{10}\sqrt{1-\epsilon_d}\right)/2=\left(1+\prod_{d=1}^{10}
e^{\mu(\cos(\Delta/2^{d-1})-1)}\right)/2.$

In Fig.~\ref{fig:rate}, we plot the key rates for $e_{\U{bit}}=0.03$, $L=32$ and $\Delta=0.2$~rad 
for the cases of $l_c=0,1,2,10$ from top to bottom. 
The top line is the key rate with no pulse correlation (i.e., $l_c=0$) that corresponds to $\Delta=0$ in Eq.~(\ref{eq:nn}).
The key rates are optimized over mean photon number $\mu$ for each value of channel transmission $\eta$. 
From these lines, we see that the pulse correlation slightly degrades the key rate 
(about 0.7 times lower than the one without pulse correlation), but the three lines with $l_c=1,2$ and 10 are almost superposed.
This implies that when the pulse correlation gets weaker as the pulses are farther apart,
which is assumed in our simulation, the long-range pulse correlation does not cause a significant impact on the key rate.

\section{Discussion}
\label{sec:dis}
In this paper, we have provided the information theoretic 
security proof of the RRDPS protocol with the pulse correlation in Alice's source by using the reference technique.
The pulse correlation is one of the serious imperfections in high-speed QKD systems
where Alice's random bit choice is propagated to the subsequent emitted pulses.
Once the number of propagated pulses ($l_c$) is fixed,
our security proof only requires the two experimentally simple assumptions on the
source:
the lower bound on the fidelity between the two $k^{\U{th}}$ states when the
correlation patterns are different
and the lower bounds on the vacuum emission probabilities of each emitted pulse.
Our numerical simulations have shown the key rates up to $l_c=10$ and have revealed
that the long-range pulse correlation does not cause a significant impact on the key rate in a realistic experimental setting.
Therefore, our security proof is effective and applicable to wide range of practical sources,
and thus paves the way to realize the truly secure and high-speed QKD systems.

We end with some open questions.
It has an practical importance to simulate the key rates based on
another source correlation model such as an intensity correlation
that is beyond the one we have supposed in our simulation shown in
Fig.~\ref{fig:rate}. 
Also, it is interesting to extend our security proof without the modification of the protocol, namely, 
with a single variable-delay interferometer assuming the same source correlation. 
Another interesting topic is to extend the security proof to accommodate quantum correlations among the emitted signals.

\section*{Acknowledgements}
This work is supported in part by the JSPS Grant-in-Aid 
for Scientific Research (C) No. 20K03779, (C) No. 21K03388, JST
Moonshot R\&D-MILLENNIA Program (grant number JPMJMS2061), JSPS KAKENHI Research (S) (Grants No: JP18H05237), 
and by CREST (Japan Science and Technology Agency) Grant No: JPMJCR1671.
AM is supported by JST, ACT-X Grant Number JPMJAX210O, Japan. 

\appendix

\section{Proof of Eq.~(\ref{eq:onlyvac})}
\label{sec:Appe1}
In this appendix, we prove Eq.~(\ref{eq:onlyvac}). 
For this, once we obtain the following proposition, by substituting $\ket{\psi}=\ket{\psi_{j_{\zeta}|j_{\zeta-1},...,j_{k+1},j_k=1,j_{k-1},...,j_1}}$, 
$\ket{\phi}=\ket{\psi_{j_{\zeta}|j_{\zeta-1},...,j_{k+1},j_k=0,j_{k-1},...,j_1}}$ and the lower bounds in Eq.~(\ref{secII:vacP}) 
to Eq.~(\ref{eq:exp}), Eq.~(\ref{eq:onlyvac}) can be obtained.
\begin{pro}
For any state $\ket{\psi}$ and $\ket{\phi}$, a lower bound on the fidelity between these two states is given by
\begin{align}
|\expect{\psi|\phi}|\ge
\begin{cases}
2\sqrt{p_{\vac,\phi}p_{\vac,\psi}}-1 & \U{if}~2\sqrt{p_{\vac,\phi}p_{\vac,\psi}}
\ge1\\
0 & \U{otherwise},
\end{cases}
\label{eq:exp}
\end{align}
where $p_{\vac,\phi}:=\tr[\ket{\vac}\bra{\vac}\ket{\phi}\bra{\phi}]$.
\end{pro}
(Proof) We expand $\ket{\psi}$ and $\ket{\phi}$ using the photon number states in
all the optical modes,
$\ket{\vac}$ and $\{\ket{n}\}_{n\ge1}$, as follows:
\begin{align}
\ket{\psi}&=\sqrt{p_{\vac,\psi}}\ket{\vac}+\sum_{n\ge1}\beta_n\ket{n},\no\\
\ket{\phi}&=\sqrt{p_{\vac,\phi}}\ket{\vac}+\sum_{n\ge1}\gamma_n\ket{n}.\no
\end{align}
We here choose the global phase of $\ket{\psi}$ and $\ket{\phi}$ such that the
coefficients of $\ket{\vac}$ being positive, and
$\beta_n\in\mathbb{C}$ and $\gamma_n\in\mathbb{C}$ are the coefficients for
$n\ge1$ of $\ket{\psi}$ and $\ket{\phi}$, respectively.
By using this, $|\expect{\psi|\phi}|$ is written as
\begin{align}
|\expect{\psi|\phi}|=\left|\sqrt{p_{\vac,\psi}p_{\vac,\phi}}+e^{\U{i}\theta}
\left|\sum_{n\ge1}\beta_n^{\ast}\gamma_n\right|\right|
\label{eq:expeq},
\end{align}
where $\theta=\arg(\sum_{n\ge1}\beta_n^{\ast}\gamma_n)$.
We next derive the upper bound on $\left|\sum_{n\ge1}\beta_n^{\ast}\gamma_n\right|$ 
by exploiting the triangle inequality and the Cauchy-Schwarz inequality:
\begin{align}
\left|\sum_{n\ge1}\beta_n^{\ast}\gamma_n\right|
&\le\sum_{n\ge1}\left|
\beta_n\right|\left|\gamma_n\right|
\le\sqrt{\left(\sum_{n\ge1}|\beta_n|^2\right)\left(\sum_{n\ge1}|\gamma_n|
^2\right)}\notag\\
&=\sqrt{(1-p_{\vac,\psi})(1-p_{\vac,\phi})}=:\tau.
\no
\end{align}
\begin{enumerate}[label=(\roman*)]
\item
If $2\sqrt{p_{\vac,\phi}p_{\vac,\psi}}\ge1$, since
$\sqrt{p_{\vac,\phi}p_{\vac,\psi}}\ge\tau$ holds, Eq.~(\ref{eq:expeq}) is lower-bounded as follows:
\begin{align}
&|\expect{\psi|\phi}|\notag\\
\ge&\left|\sqrt{p_{\vac,\phi}p_{\vac,\psi}}+e^{\U{i}\pi}\tau\right|\no\\
=&\sqrt{p_{\vac,\phi}p_{\vac,\psi}}-\sqrt{(1-p_{\vac,\psi})(1-p_{\vac,\phi})}\no\\
\ge&\sqrt{p_{\vac,\phi}p_{\vac,\psi}}-\sqrt{1+p_{\vac,\psi}p_{\vac,\phi}-2\sqrt{p_{\vac,\phi}p_{\vac,\psi}}}\no\\
=&2\sqrt{p_{\vac,\phi}p_{\vac,\psi}}-1
\no.
\end{align}
The second inequality follows from the fact that $a+b\ge2\sqrt{ab}$ holds for any $a,b\ge0$.
\item
If $2\sqrt{p_{\vac,\phi}p_{\vac,\psi}}<1$, we only have the trivial lower bound:
\begin{align}
|\expect{\psi|\phi}|\ge0.
\end{align}
\end{enumerate}

\section{Proof of Eqs.~(\ref{mainstate}) and (\ref{eq:lowa})}
\label{sec:ANYCOMP}
In this appendix, we prove Eqs. (\ref{mainstate}) and (\ref{eq:lowa}). We start form Eq.~(\ref{maink-1}):
\begin{widetext}
\begin{align}
\ket{\psi^{\U{Act}}_{j_t|\bm{j}_{t-1}}}_{\bm{A}_{\ge t+1}\bm{B}_{\ge t}}
=e^{\U{i}\theta_{j_t|\bm{j}_{t-1}}}\ket{\psi_{j_t|\bm{j}_{t-1}}}_{B_t}
\left(
\frac{1}{\sqrt{2^{N-t}}}
\sum_{j_N}\cdots\sum_{j_{t+1}}\bigotimes_{\zeta=t+1}^N\ket{j_{\zeta}}_{A_{\zeta}}
e^{\U{i}\theta_{j_\zeta|\bm{j}_{\zeta-1}}}\ket{\psi_{j_{\zeta}|j_{\zeta-1},...,j_{t+1},j_t,
\bm{j}_{t-1}}}_{B_{\zeta}}
\right).
\label{Scha}
\end{align}
\end{widetext}
To see how the information $j_t$ is encoded to the state $\ket{\psi^{\U{Act}}_{j_t|\bm{j}_{t-1}}}_{\bm{A}_{\ge t+1}\bm{B}_{\ge t}}$, 
we expand it using $\ket{\Phi_{\bm{j}_{t-1}}}_{\bm{A}_{\ge t+1}\bm{B}_{\ge t+1}}$ and
$\ket{\Phi^{\perp}_{j_t,\bm{j}_{t-1}}}_{\bm{A}_{\ge t+1}\bm{B}_{\ge t+1}}$ to have
\begin{align}
&\ket{\psi^{\U{Act}}_{j_t|\bm{j}_{t-1}}}_{\bm{A}_{\ge t+1}\bm{B}_{\ge t}}
=e^{\U{i}\theta_{j_t|\bm{j}_{t-1}}}\ket{\psi_{j_t|\bm{j}_{t-1}}}_{B_t}
\otimes\no\\
&\left(a_{j_t,\bm{j}_{t-1}}\ket{\Phi_{\bm{j}_{t-1}}}_{\bm{A}_{\ge t+1}\bm{B}_{\ge t+1}}+
b_{j_t,\bm{j}_{t-1}}\ket{\Phi^{\perp}_{j_t,\bm{j}_{t-1}}}_{\bm{A}_{\ge t+1}\bm{B}_{\ge
t+1}}\right),
\label{SDEC}
\end{align}
where $\ket{\Phi_{\bm{j}_{t-1}}}_{\bm{A}_{\ge t+1}\bm{B}_{\ge t+1}}$ and $
\ket{\Phi^{\perp}_{j_t,\bm{j}_{t-1}}}_{\bm{A}_{\ge t+1}\bm{B}_{\ge t+1}}$ denote
some normalized states, and these are orthogonal each other.
The subscripts in $a_{j_t,\bm{j}_{t-1}}$, $b_{j_t,\bm{j}_{t-1}}$, $\ket{\Phi_{\bm{j}_{t-1}}}
_{\bm{A}_{\ge t+1}\bm{B}_{\ge t+1}}$,
and $\ket{\Phi^{\perp}_{j_t,\bm{j}_{t-1}}}_{\bm{A}_{\ge t+1}\bm{B}_{\ge t+1}}$
indicate the dependency on the previous setting choices
\footnote{Note that these subscripts are $j_t,j_{t-1},...,j_1$
if $1\le t\le l_c+1$, and $j_t,j_{t-1},...,j_{t-l_c}$ if $l_c+2\le t$.}.
Importantly, $\ket{\Phi_{\bm{j}_{t-1}}}_{\bm{A}_{\ge t+1}\bm{B}_{\ge t+1}}$ does not
depend on $j_t$ but
$\ket{\Phi^{\perp}_{j_t,\bm{j}_{t-1}}}_{\bm{A}_{\ge t+1}\bm{B}_{\ge t+1}}$ does.
This means that $\ket{\Phi^{\perp}_{j_t,\bm{j}_{t-1}}}_{\bm{A}_{\ge t+1}\bm{B}_{\ge
t+1}}$ represents the side-channel state of $j_t$.
For $\ket{\Phi_{\bm{j}_{t-1}}}_{\bm{A}_{\ge t+1}\bm{B}_{\ge t+1}}$, we can take any
state as long as it is independent of $j_t$.
Here, we choose it as
\begin{widetext}
\begin{align}
\ket{\Phi_{\bm{j}_{t-1}}}_{\bm{A}_{\ge t+1}\bm{B}_{\ge t+1}}=\frac{1}{\sqrt{2^{N-t}}}
&
\left(\sum_{j_{t+l_c}}\cdots\sum_{j_{t+1}}\bigotimes_{\zeta=t+1}^{t+l_c}
\ket{j_\zeta}_{A_\zeta}\ket{\psi_{j_{\zeta}|j_{\zeta-1},...,j_{t+1},j_t=0,\bm{j}_{t-1}}}
_{B_{\zeta}}\right)\no\\
\otimes&
\left(\sum_{j_N}\cdots\sum_{j_{t+l_c+1}}\bigotimes_{\zeta=t+l_c+1}^{N}
e^{\U{i}\theta_{j_\zeta|\bm{j}_{\zeta-1}}}\ket{j_\zeta}_{A_\zeta}\ket{\psi_{j_{\zeta}|
j_{\zeta-1},...,j_{t+1},j_t=0,\bm{j}_{t-1}}}_{B_{\zeta}}\right)
\label{wechoose}
\end{align}
\end{widetext}
that corresponds to the $N-t$ systems of Eq.~(\ref{Scha}) with $j_t$ being fixed to 
be 0 and with omitting the phase from the state 
$\{\ket{\psi_{j_{\zeta}|j_{\zeta-1},...,j_{t+1},j_t=0,\bm{j}_{t-1}}}_{B_{\zeta}}\}_{\zeta=t+1}^{t+l_c}$.
The reason for omitting the phase is to guarantee the positivity of $a_{j_t,\bm{j}_{t-1}}$
in Eq.~(\ref{SDEC}).
The remaining task is to derive the lower bound on $a_{j_t,\bm{j}_{t-1}}$ for $j_t\in\{0,1\}$ using the assumption~(A2).
Since $a_{j_t,\bm{j}_{t-1}}$ is the inner product between Eq.~(\ref{wechoose}) and the vector
\begin{align}
&\frac{1}{\sqrt{2^{N-t}}}
\sum_{j_N}\cdots\sum_{j_{t+1}}\no\\
&\bigotimes_{\zeta=t+1}^N\ket{j_{\zeta}}_{A_{\zeta}}
e^{\U{i}\theta_{j_\zeta|\bm{j}_{\zeta-1}}}\ket{\psi_{j_{\zeta}|j_{\zeta-1},...,j_{t+1},j_t,\bm{j}_{t-1}}}_{B_{\zeta}},
\no
\end{align}
which is the state of the $N-t$ systems shown in the parenthesis of Eq.~(\ref{Scha}),
we have
\begin{widetext}
\begin{align}
a_{j_t,\bm{j}_{t-1}}&=
\frac{1}{2^{N-t}}
\left(\sum_{j_{t+l_c}}\cdots\sum_{j_{t+1}}\prod_{\zeta=t+1}^{t+l_c}
e^{\U{i}\theta_{j_\zeta|\bm{j}_{\zeta-1}}}
\expect{\psi_{j_{\zeta}|j_{\zeta-1},...,j_{t+1},j_t=0,\bm{j}_{t-1}}|\psi_{j_{\zeta}|
j_{\zeta-1},...,j_{t+1},j_t, \bm{j}_{t-1}}}
\right)
\left(\sum_{j_N}\cdots\sum_{j_{t+l_c+1}}1\right)
\notag\\
&=
\frac{1}{2^{l_c}}
\sum_{j_{t+l_c}}\cdots\sum_{j_{t+1}}\prod_{\zeta=t+1}^{t+l_c}
e^{\U{i}\theta_{j_\zeta|\bm{j}_{\zeta-1}}}
\expect{\psi_{j_{\zeta}|j_{\zeta-1},...,j_{t+1},j_t=0,\bm{j}_{t-1}}|\psi_{j_{\zeta}|
j_{\zeta-1},...,j_{t+1},j_t, \bm{j}_{t-1}}}
\no\\
&=
\frac{1}{2^{l_c}}
\sum_{j_{t+l_c}}\cdots\sum_{j_{t+1}}\prod_{\zeta=t+1}^{t+l_c}
\left|\expect{\psi_{j_{\zeta}|j_{\zeta-1},...,j_{t+1},j_t=0,\bm{j}_{t-1}}|\psi_{j_{\zeta}|j_{\zeta-1},...,j_{t+1},j_t, \bm{j}_{t-1}}}\right|.
\label{aabs}
\end{align}
\end{widetext}
In the second equality, we set the phases $e^{\U{i}\theta_{j_\zeta|\bm{j}_{\zeta-1}}}$ 
for any $\zeta$ ($t+1\le\zeta\le t+l_c$) and $\bm{j}_N$ as
\begin{align}
&e^{\U{i}\theta_{j_\zeta|\bm{j}_{\zeta-1}}}\no\\
&:=
\frac{\left|\expect{\psi_{j_\zeta|j_{\zeta-1},...,j_{t+1},j_t=0,\bm{j}_{t-1}}|\psi_{j_\zeta|j_{\zeta-1},...,j_{t+1},j_t,\bm{j}_{t-1}}}\right|}
{\expect{\psi_{j_\zeta|j_{\zeta-1},...,j_{t+1},j_t=0,\bm{j}_{t-1}}|\psi_{j_\zeta|j_{\zeta-1},...,j_{t+1},j_t,\bm{j}_{t-1}}}}.
\label{ephaseA}
\end{align}
Since the only difference between both states in the inner product of Eq.~(\ref{aabs}) is in the $j_t^{\U{th}}$ index, we have
$$
a_{j_t=0,\bm{j}_{t-1}}=1.
$$
On the other hand, if $j_t=1$, by applying Eq.~(\ref{secII:LRASS1}) to Eq.~(\ref{aabs}), we obtain
\begin{align}
a_{j_t=1,\bm{j}_{t-1}} & \ge\frac{1}{2^{l_c}}\sum_{j_{t+l_c}}\cdots\sum_{j_{t+1}}\prod_{\zeta=t+1}^{t+l_c}\sqrt{1-\epsilon_{\zeta-t}}
\no\\
&=\prod_{d=1}^{l_c}\sqrt{1-\epsilon_{d}}.\no
\end{align}
This ends the proof of Eqs.~(\ref{mainstate}) and (\ref{eq:lowa}). \sq

\section{Proof of Eq.~(\ref{eq:keyrate})}
\label{ap:all}
In this appendix, we prove Eq.~(\ref{eq:keyrate}) that is our result of the security proof. 
Our security proof adopts the composability definition~\cite{composable}, where 
the security of our RRDPS protocol is evaluated by the correctness and secrecy
parameters. 
As shown in~\cite{koashi2009}, these parameters can be quantified separately, and
since the correctness parameter is obtained by a verification step of the protocol, our
target is to compute the secrecy parameter $\epsilon_s$.
The protocol is $\epsilon_s$-secret if and only if
\begin{align}
d:=||\rho^{\fin}_{\bm{A}_{\fin}E}-\rho^{\id}_{\bm{A}_{\fin}E}||
\le \epsilon_s.
\label{def:tra}
\end{align}
Here, we define trace distance $||X||:=\tr[\sqrt{X^\dagger X}]/2$,
$\rho^{\fin}_{\bm{A}_{\fin}E}$ denotes the state of Alice's actual final keys
and Eve's quantum system, and $\rho^{\id}_{\bm{A}_{\fin}E}$ denotes the
state of ideal final keys that are completely secret from Eve and Eve's quantum system.
These final keys can be obtained by applying a quantum circuit (composed of a lot of
CNOT gates), which is determined by random matrices used in privacy amplification.
We introduce a quantum operation $\mathcal{E}^{(w)}_{\U{act}}$ that
extracts the $w^{\U{th}}$-type final key from $w^{\U{th}}$-type sifted qubits of
systems $\bm{A}_{\sif}^{(w)}$.
The operation $\mathcal{E}^{(w)}_{\U{act}}$ is composed of CNOT gates acting
on systems $\bm{A}^{(w)}_{\sif}$ and $Z$-basis measurements.
The total operation in privacy amplification to obtain the final keys, 
which acts on all the sifted qubits of systems 
$\bm{A}_{\sif}:=\bm{A}^{(1)}_{\sif}\bm{A}^{(2)}_{\sif}...\bm{A}^{(l_c+1)}_{\sif}$, 
is then written as
\begin{align}
\mathcal{E}_{\U{act}}:=\bigotimes_{w=1}^{l_c+1}\mathcal{E}^{(w)}_{\U{act}}.
\label{Ea}
\end{align}
Using this definition, $\rho^{\fin}_{\bm{A}_{\fin}E}$ and $\rho^{\id}_{\bm{A}_{\fin}E}$
in Eq.~(\ref{def:tra}) can be
written as
\begin{align}
&\rho^{\fin}_{\bm{A}_{\fin}E}=\mathcal{E}_{\U{act}}(\rho_{\bm{A}_{\sif}E}),\label{decomp2}\\
&\rho^{\id}_{\bm{A}_{\fin}E}=\mathcal{E}_{\U{act}}(\ket{\bm{+}}\bra{\bm{+}}_{\bm{A}_{\sif}}\otimes \tr_{\bm{A}
_{\sif}}[\rho_{\bm{A}_{\sif}E}]),
\label{decomp}
\end{align}
where $\rho_{\bm{A}_{\sif}E}$ is the state of Alice's all the sifted qubits and Eve's
quantum system just before
executing privacy amplification $\mathcal{E}_{\U{act}}$.
Note that $\ket{+}:=(\ket{0}+\ket{1})/\sqrt{2}$ is the $X$-basis eigenstate from which
an ideal key
can be extracted. We use the notation $\ket{\bm{+}}_{\bm{A}_{\sif}}$ to express all
the qubits of systems $\bm{A}_{\sif}$ being in $\ket{+}$.
Below, we show that the above trace distance $d$ can be upper-bounded by using our
Theorem~\ref{th1}. 
In this theorem, 
as we consider the asymptotic limit of an infinite sifted key length, we neglect the
probability of failing in obtaining the upper bound of Eq.~(\ref{eq:resulteph}).
For a general discussion here, we denote its negligible
probability of failing in obtaining Eq.~(\ref{eq:resulteph}) for $w$ by $\xi^{(w)}$.
With these failure probabilities and Theorem~\ref{th1}, we have the following
proposition.
\begin{pro}
When Eq.~(\ref{eq:resulteph}) in our Theorem~\ref{th1} holds except for probability $\xi^{(w)}$, and
if the amount of privacy amplification applied for the $w^{\U{th}}$-type reconciled key is set to be
\begin{align}
N^{(w)}_{\U{suc}}h(e^{(w),\U{U}}_{\U{ph}})+\log_2\frac{1}{\eta^{(w)}}
\label{result:phw}
\end{align}
for any $\eta^{(w)}>0$,
the secrecy parameter $\epsilon_s$ of the RRDPS protocol is given by
\begin{align}
\epsilon_s=\sum_{w=1}^{l_c+1}\sqrt{2}\sqrt{\xi^{(w)}+\eta^{(w)}}.
\label{pro}
\end{align}
Here, $h(x)$ denotes the binary entropy function, and $N^{(w)}_{\U{suc}}$ is the
number of sifted qubits of systems $\bm{A}^{(w)}_{\sif}$.
\end{pro}
In the asymptotic limit ($N^{(w)}_{\U{suc}}\to\infty$), as $\xi^{(w)}\to0$ and $
\eta^{(w)}\to0$,
$\epsilon_s$ in Eq.~(\ref{pro}) results in negligible.
Therefore, using this proposition by setting $\xi^{(w)}\to0$ and $\eta^{(w)}\to0$,
we finally obtain the secret key rate shown in Eq.~(\ref{eq:keyrate}) with $\epsilon_s$-secret.
The rest of this appendix is devoted to proving this proposition.
\\\\
(Proof)
First, when the amount of privacy amplification is set to be as Eq.~(\ref{result:phw}), it is straightforward from~\cite{HayashiNaka} to
derive the secrecy parameter for the $w^{\U{th}}$-type key, that is,
we obtain for any $w\in\{1,2,...,l_c+1\}$,
\begin{align}
&||\mathcal{E}^{(\ge w)}_{\U{act}}(\tr_{\bm{A}^{(0,...,w-1)}_{\sif}}[\rho_{\bm{A}_{\sif}E}])\no\\
&-\mathcal{E}^{(\ge w)}_{\U{act}}(\ket{\bm{+}}\bra{\bm{+}}_{\bm{A}^{(w)}_{\sif}}\otimes
\tr_{\bm{A}^{(1,...,w)}_{\sif}}[\rho_{\bm{A}_{\sif}E}])||\no\\
&\le\sqrt{2}\sqrt{\xi^{(w)}+\eta^{(w)}}=:\Delta^{(w)}.
\label{eaches}
\end{align}
Here, we define $\mathcal{E}^{(\ge w)}_{\U{act}}:=\bigotimes_{x=w}^{l_c+1}\mathcal{E}^{(x)}_{\U{act}}$ and 
$\tr_{\bm{A}^{(0)}_{\sif}}$ means that no system is traced out. 
This bound $\Delta^{(w)}$ is obtained by executing 
phase error correction to correct all the qubits of systems $\bm{A}^{(w)}_{\sif}$ to $\ket{+}$. 
This operation of its correction does not change any statistics of the measurement outcomes obtained by 
$\mathcal{E}^{(\ge w)}_{\U{act}}$, 
and hence we can insert this operation to upper-bound the trace distance in Eq.~(\ref{eaches}). 
Then, based on~\cite{HayashiNaka}, 
this trace distance can be evaluated by 
the failure probability $\eta^{(w)}$ of phase error correction when $e^{(w),\U{U}}_{\U{ph}}$ is obtained and 
the failure probability $\xi^{(w)}$ of obtaining the upper-bound on the phase error rate 
$e^{(w),\U{U}}_{\U{ph}}$. 
We remark that in doing this argument of phase error correction, 
the state $\rho_{\bm{A}_{\sif}E}$ must be dependent on $w\in\{1,2,...,l_c+1\}$. 
This is 
because we define the virtual state $\ket{\Psi}_{\bm{A}_N\bm{B}_N}$ in Eq.~(\ref{eq:entbased})
for each $w$ differently (due to the phase factors $e^{\U{i}\theta_{j_k|\bm{j}_{k-1}}}$ defined in Eqs.~(\ref{eq:main_ph_1}) and (\ref{eq:main_ph}))
in order to obtain the upper bound on the phase error rate for each $w$, which is explained in Sec.~\ref{app:dec}. 
The differences of the states $\rho_{\bm{A}_{\sif}E}$ for $w$
become apparent in correcting the phase errors for each $w^{\U{th}}$-type sifted qubits. 
Importantly, however,
these differences do not change any statistics of the final keys obtained through
the operation $\mathcal{E}_{\U{act}}$
\footnote{
In other words, the unitary operator acting on Alice's virtual qubits of systems $
\bm{A}_N$ to add the adequate
phase factors for each $w$ commutes with $\mathcal{E}^{(w)}_{\U{act}}$ since this unitary
operator is diagonal in the $Z$-basis.
}.
Therefore, we can take the state $\rho_{\bm{A}_{\sif}E}$ independently of $w$ 
when we consider the security of the final keys in Eq.~(\ref{eaches}). 
This is the reason why the state $\rho_{\bm{A}_{\sif}E}$ in Eq.~(\ref{eaches}) does not depend on $w$.

Then, by exploiting Eq.~(\ref{eaches}) and substituting Eqs.~(\ref{decomp2}) and (\ref{decomp}) to
Eq.~(\ref{def:tra}), $d$ is calculated as follows:
\begin{widetext}
\begin{align}
d&\le
||\mathcal{E}_{\U{act}}(\rho_{\bm{A}_{\sif}E})-\mathcal{E}_{\U{act}}(\ket{\bm{+}}
\bra{\bm{+}}_{\bm{A}^{(w=1)}_{\sif}}
\otimes\tr_{\bm{A}^{(w=1)}_\sif}[\rho_{\bm{A}_\sif E}])||\no\\
&+
||\mathcal{E}_{\U{act}}(\ket{\bm{+}}\bra{\bm{+}}_{\bm{A}^{(w=1)}_{\sif}}
\otimes\tr_{\bm{A}^{(w=1)}_\sif}[\rho_{\bm{A}_\sif E}])-
\mathcal{E}_{\U{act}}(\ket{\bm{+}}\bra{\bm{+}}_{\bm{A}_{\sif}}\otimes \tr_{\bm{A}
_{\sif}}[\rho_{\bm{A}_{\sif}E}])||\\
&\le\Delta^{(1)}+
||\mathcal{E}^{(\ge2)}_{\U{act}}(\tr_{\bm{A}^{(w=1)}_\sif}[\rho_{\bm{A}_\sif E}])-
\mathcal{E}^{(\ge2)}_{\U{act}}(\ket{\bm{+}}\bra{\bm{+}}_{\bm{A}^{(w\ge2)}_{\sif}}
\otimes \tr_{\bm{A}_{\sif}}[\rho_{\bm{A}_{\sif}E}])||\\
&\le\Delta^{(1)}+
||\mathcal{E}^{(\ge2)}_{\U{act}}(\tr_{\bm{A}^{(w=1)}_\sif}[\rho_{\bm{A}_\sif E}])-
\mathcal{E}^{(\ge2)}_{\U{act}}(\ket{\bm{+}}\bra{\bm{+}}_{\bm{A}^{(w=2)}_{\sif}}
\otimes \tr_{\bm{A}^{(w=1,2)}_{\sif}}[\rho_{\bm{A}_{\sif}E}])||\no\\
&+||\mathcal{E}^{(\ge2)}_{\U{act}}(\ket{\bm{+}}\bra{\bm{+}}_{\bm{A}^{(w=2)}
_{\sif}}\otimes \tr_{\bm{A}^{(w=1,2)}_{\sif}}[\rho_{\bm{A}_{\sif}E}])
-\mathcal{E}^{(\ge2)}_{\U{act}}(\ket{\bm{+}}\bra{\bm{+}}_{\bm{A}^{(w\ge2)}_{\sif}}
\otimes \tr_{\bm{A}_{\sif}}[\rho_{\bm{A}_{\sif}E}])||\\
&\le\Delta^{(1)}+\Delta^{(2)}
+||\mathcal{E}^{(\ge2)}_{\U{act}}(\ket{\bm{+}}\bra{\bm{+}}_{\bm{A}^{(w=2)}_{\sif}}
\otimes \tr_{\bm{A}^{(w=1,2)}_{\sif}}[\rho_{\bm{A}_{\sif}E}])
-\mathcal{E}^{(\ge2)}_{\U{act}}(\ket{\bm{+}}\bra{\bm{+}}_{\bm{A}^{(w\ge2)}_{\sif}}
\otimes \tr_{\bm{A}_{\sif}}[\rho_{\bm{A}_{\sif}E}])||.
\end{align}
\end{widetext}
The first and third inequalities come from the triangle inequality of trace distance, and the
second and fourth ones follow from Eq.~(\ref{eaches}). 
So far, we have quantified the security of the $w^{\U{th}}$-type keys for $w=1,2$.
By repeating the same arguments for $w=3,4,...,l_c$, we have
\begin{widetext}
\begin{align}
d\le \sum_{w=1}^{l_c}\Delta^{(w)}
+||\mathcal{E}^{(\ge l_c+1)}_{\U{act}}(\tr_{\bm{A}^{(w=1,2...,l_c)}_\sif}[\rho_{\bm{A}
_\sif E}])-
\mathcal{E}^{(\ge l_c+1)}_{\U{act}}(\ket{\bm{+}}\bra{\bm{+}}_{\bm{A}^{(w=l_c+1)}
_{\sif}}\otimes \tr_{\bm{A}_{\sif}}[\rho_{\bm{A}_{\sif}E}])||.
\end{align}
\end{widetext}
Finally, applying Eq.~(\ref{eaches}) for $w=l_c+1$, we obtain
\begin{align}
d&\le \sum_{w=1}^{l_c+1}\Delta^{(w)},
\end{align}
which ends the proof.
\sq

\end{document}